\documentclass[10pt,journal]{IEEEtran}
\pdfoutput=1
\usepackage{amsfonts}
\usepackage[mathcal]{euscript} 
\usepackage[cmex10]{amsmath}
\usepackage{amssymb}
\usepackage{graphics,graphicx}
\usepackage{cite,calc}
\usepackage{subdepth}
\usepackage{upgreek}

\usepackage{hyperref}
\hypersetup{colorlinks = true, citecolor = blue}

\usepackage{color}
\usepackage{enumerate}
\usepackage[english]{babel} 
\usepackage{url}

\usepackage{enumitem}
\usepackage{tabularx}

\usepackage{stackengine}
\newcommand\xrowht[2][0]{\addstackgap[1\dimexpr#2\relax]{\vphantom{#1}}}

\usepackage{float}
\usepackage{array, makecell}
\setcellgapes{5pt}

\newtheorem{definition}{Definition}
\newtheorem{theorem}{Theorem}
\newtheorem{lemma}[theorem]{Lemma}

\newtheorem{corollary}[theorem]{Corollary}

\usepackage[thinlines]{easytable}
\newcolumntype{M}[1]{>{\centering\arraybackslash}m{#1}}
\newcolumntype{N}{@{}m{0pt}@{}}

{\end{list}}
\DeclareMathOperator*{\argmax}{arg\,max}
\DeclareMathOperator*{\argmin}{arg\,min}
\DeclareMathAlphabet{\mathbbb}{U}{bbold}{m}{n}  
\newcommand{\p}{\partial}
\newcommand{\ds}{\displaystyle}
\newcommand{\ts}{\textstyle}
\newcommand{\kron}{\mathbbb{1}}
\newcommand{\reals}{\mathbb{R}}
\newcommand{\defeq}{\triangleq}
\newcommand{\gwomit}[1]{}
\newcommand{\secref}[1]{Section~\ref{#1}}
\newcommand{\defref}[1]{Definition~\ref{#1}}
\newcommand{\lemref}[1]{Lemma~\ref{#1}}
\newcommand{\thmref}[1]{Theorem~\ref{#1}}
\newcommand{\appref}[1]{Appendix~\ref{#1}}

\newcommand{\corolref}[1]{Corollary~\ref{#1}}
\newcommand{\figref}[1]{Fig.~\ref{#1}}
\newcommand{\tabref}[1]{Table~\ref{#1}}
\newcommand{\ev}{{\mathbb{E}}}

\newcommand{\E}[1]{\ev\left[#1\right]}

\newcommand{\Ed}[2]{\ev_{#1}\left[#2\right]}

\newcommand{\cC}{\mathcal{C}}

\newcommand{\cH}{\mathcal{H}}
\newcommand{\cI}{\mathcal{I}}

\newcommand{\cP}{\mathcal{P}}
\newcommand{\cS}{\mathcal{S}}
\newcommand{\cT}{\mathcal{T}}

\newcommand{\cY}{\mathcal{Y}}
\newcommand{\diff}{\mathrm{d}}
\newcommand{\ip}[2]{\langle#1,#2\rangle}
\newcommand{\T}{\mathrm{T}}
\newcommand{\markov}{\leftrightarrow}
\newcommand{\e}{\mathrm{e}}

\newcommand{\yh}{\hat{y}}
\newcommand{\Vt}{\tilde{V}}
\def\setTh{\text{$\mksetTh$}}  
\def\mksetTh{\setbox0\hbox{$\mathcal{O}$}%
    \rlap{\hbox to \wd0{\hss\raisebox{.2ex}{\hspace*{.07em}\textbf{-}}\hss}}\box0}

\begin{document}
\title{Bregman Divergence Bounds and \\ Universality Properties of the
  Logarithmic Loss} 

\author{Amichai~Painsky,~\IEEEmembership{Member,~IEEE,}
        and Gregory~W.~Wornell,~\IEEEmembership{Fellow,~IEEE}%
\thanks{Manuscript received Oct.\ 2018, revised July 2019.
The material in this paper was presented in part at the
  Int.\ Symp.\ Inform.\ Theory (ISIT-2018)
  \cite{painsky2018universality}, Vail, Colorado, June 2018.} 
\thanks{This work was supported, in part, by NSF under Grant No.~CCF-1717610.}
\thanks{A.~Painsky is with the Department of Industrial Engineering,
  Tel Aviv University, Tel Aviv 6139001, Israel  (E-mail:
  amichaip@tauex.tau.ac.il).  G.~W.~Wornell is with the Department of
  Electrical Engineering  and Computer Science, Massachusetts
  Institute of Technology, Cambridge, MA 02139 (E-mail: gww@mit.edu).} 
}

\maketitle

\begin{abstract}
A loss function measures the discrepancy between the true values and
their estimated fits, for a given instance of data. In classification
problems, a loss function is said to be proper if a minimizer of the
expected loss is the true underlying probability. We show that for
binary classification, the divergence associated with smooth, proper,
and convex loss functions is upper bounded by the Kullback-Leibler
(KL) divergence, to within a normalization constant. This implies that by
minimizing the logarithmic loss associated with the KL divergence, we
minimize an upper bound to any choice of loss from this set.  As such
the logarithmic loss is universal in the sense of providing
performance guarantees with respect to a broad class of accuracy
measures. Importantly, this notion of universality is not
problem-specific, enabling its use in diverse applications, including
predictive modeling, data clustering and sample complexity
analysis. Generalizations to arbitary finite alphabets are also
developed.  The derived inequalities extend several well-known
$f$-divergence results.
\end{abstract}

.

\section{Introduction}
\label{intro}

\IEEEPARstart{O}{ne} of the major roles of statistical analysis is
making predictions about future events and providing suitable
accuracy guarantees.  For example, consider a weather forecaster
that estimates the chances of rain on the following day.  Its
performance may be evaluated by multiple statistical measures.  We may
count the number of times it assessed the chance of rain as greater
than 50\%, when there was eventually no rain (and vice versa).  This
corresponds to the so-called 0-1 loss, with threshold parameter $t=1/2$.
Alternatively, we may consider a variety of values for $t$, or even a
completely different measure.  Indeed, there are many candidates,
including the quadratic loss, Bernoulli log-likelihood loss, boosting
loss, etc.~\cite{buja2005loss}.  Choosing a good measure is a
well-studied problem, mostly in the context of \emph{scoring rules} in
decision theory
\cite{winkler1996scoring,gneiting2007strictly,merkle2013choosing,dawid2014theory}.

Assuming that the desired measure is known in advance, the predictor
may be designed accordingly---i.e., to minimize that measure.
However, in practice, different tasks may require inferring different
information from the provided estimates.  Moreover, designing a
predictor with respect to one measure may result in poor performance
when evaluated by another.  For example, the minimizer of a 0-1 loss
may result in an unbounded loss, when measured with a Bernoulli
log-likelihood loss.  In such cases, it would be desireable to design
a predictor according to a ``universal'' measure, i.e., one that is
suitable for a variety of purposes, and provide performance guarantees
for different uses.  

In this paper, we show that for binary classification, the Bernoulli
log-likelihood loss (log-loss) is such a universal choice, dominating
all alternative ``analytically convenient'' (i.e., smooth, proper, and
convex) loss functions.  Specifically, we show that by minimizing the
log-loss we minimize the regret associated with all possible
alternatives from this set.  Our result justifies the use of log-loss
in many applications.

As we develop, our universality result may be equivalently viewed from
a divergence analysis viewpoint. In particular, we establish that the
divergence associated with the log-loss---i.e., Kullback Leibler (KL)
divergence---upper bounds a set of Bregman divergences that satisfy a
condition on its Hessian.  Additionally, we show that any separable
Bregman divergence that is convex in its second argument is a member
of this set.  This result provides a new set of Bregman divergence
inequalities.  In this sense, our Bregman analysis is complementary to
the well-known $f$-divergence inequality results
\cite{sason2018f,sason2016f, harremoes2011pairs,reid2011information}.

We further develop several applications for our results, including
universal forecasting, universal data clustering, and universal sample
complexity analysis for learning problems, in addition to establishing
the universality of the information bottleneck principle.  We
emphasize that our universality results are derived in a rather
general setting, and not restricted to a specific problem.  As such,
they may find a wide range of additional applications.

The remainder of the paper is organized as follows.
\secref{Previous Work} summarizes related work on loss function
analysis, universality and divergence inequalities.   \secref{basic}
provides the needed notation, terminology, and definitions.
\secref{main result} contains the main results for binary alphabets,
and their generalization to arbitrary finite alphabets is developed in
\secref{Bregman_Divergence_Inequalities}.   Additional numerical
analysis and experimental validation is provided in
\secref{illustrations}, and the implications of our results 
in three distinct applications is described in \secref{application}.
Finally, \secref{conc} contains some concluding remarks.

\section{Related Work}
\label{Previous Work}

The Bernoulli log-likelihood loss function plays a fundamental role in
information theory, machine learning, statistics and many other
disciplines.  Its unique properties and broad applications have been
extensively studied over the years. 

The Bernoulli log-likelihood loss function arises naturally in the
context of parameter estimation.  Consider a set of independent,
identically distributed (i.i.d.) observations $y^n=(y_1,\dots,y_n)$
drawn from a distribution $p_Y(\cdot;\theta)$ whose parameter $\theta$
is unknown.  Then the maximum likelihood estimate of $\theta$ in
$\setTh$ is
\begin{equation*} 
\hat{\theta}=\argmax_{\theta \in \setTh}
L(\theta;y^n),
\end{equation*}
where
\begin{equation*}
L(\theta;y^n)=p_{Y^n}(y^n;\theta)=\prod_{i=1}^n p_Y(y_i;\theta).
\end{equation*}
Intuitively, it selects the parameters values that make the data most
probable.  Equivalently, this estimate minimizes a loss that is the
(negative, normalized, natural) logarithm of the likelihood function,
viz.,
\begin{equation*} 
  \ell (\theta ;y^n) \defeq -\frac1n \log L(\theta;y^n) 
= -\frac1n \sum_{i=1}^n \log p_Y(y_i;\theta), 
\end{equation*}
whose mean is 
\begin{equation*} 
\E{\ell(\theta;Y)} = -\E{\log p_Y(Y;\theta)}.
\end{equation*}
Hence, by minimizing this Bernoulli log-likelihood loss, termed the
\emph{log-loss}, over a set of parameters we maximize the likelihood
of the given observations.

The log-loss also arises naturally in information theory.  The
\emph{self-information} loss function $-\log p_Y(y)$ defines the ideal
codeword length for describing the realization $Y=y$
\cite{cover2012elements}.  In this sense, minimizing the log-loss
corresponds to minimizing the amount of information that are necessary
to convey the observed realizations.  Further, the expected
self-information is simply Shannon's entropy which reflects the
average uncertainty associated with sampling the random variable $Y$.

The logarithmic loss function is known to be ``universal'' from several
information-theoretic points of view.   In \cite{merhav1998universal},
Feder and Merhav consider the problem of universal sequential
prediction, where a future observation is to be estimated from a given
set of past observations.  The notion of universality comes from the
assumption that the underlaying distribution is unknown, or even
nonexistent.  In this case, it is shown that if there exists a
universal predictor (with a uniformly rapidly decaying redundancy
rates) that minimizes the logarithmic loss function, then there exist
universal predictors for any other loss function.

More recently, No and Weissman \cite{no2015universality} introduced
log-loss universality results in the context of lossy
compression.  They show that for any fixed length lossy compression
problem under an arbitrary distortion criterion, there is an
equivalent lossy compression problem under a log-loss criterion where
the optimum schemes coincide.   This result implies that without loss
of generality, one may restrict attention to the log-loss problem
(under an appropriate reconstruction alphabet).   In addition,
\cite{no2015universality} considers the successive refinement problem,
showing that if the first decoder operates under log-loss, then any
discrete memoryless source is successively refinable under an
arbitrary distortion criterion for the second decoder.

It is important to emphasize that universality results of the type
discussed above are largely limited to relatively narrowly-defined
problems and specific optimization criteria.  By contrast, our
development is aimed at a broader notion of universality that is not
restricted to a specific problem, and considers a broader range of
criteria.

An additional information-theoretic justification for the wide use of
the log-loss is introduced in \cite{jiao2015justification}.  This work
focuses on statistical inference with side information, showing that
for an alphabet size greater than two, the  log-loss is the only loss
function that benefits from side information and satisfies the data
processing lemma.  This result extends some well-known properties of
the log-loss with respect to the data processing lemma, as later
described. 

Within decision theory, statistical learning and inference problems,
the log-loss also plays further key role in the context of
\emph{proper} loss function, which produce estimates that are unbiased
with respect to the true underlaying distribution.  Proper loss
functions have been extensively studied, compared, and suggested for a
variety of tasks \cite{winkler1996scoring, gneiting2007strictly,
  bickel2007some, merkle2013choosing, dawid2014theory}.
Among these, the log-loss is
special: it is the only proper loss that is \emph{local}
\cite{bs00,parry2012proper}.  This means that the log-loss is the only
proper loss function that assigns an estimate for the probability of
the event $Y=y_0$ that depends only on the outcome $Y=y_0$.  

In turn, proper loss functions are closely related to Bregman
divergences, with which there exists a one-to-one correspondence
\cite{gneiting2007strictly}.  For the log-loss, the associated Bregman
divergence is KL divergence, which is also an
instance of an $f$-divergence \cite{cs04}.  Significantly, for
probability distributions, the KL divergence is the only divergence
measure that is a member of both of these classes of divergences
\cite{csiszar1995generalized}.  The Bregman divergences are the only
divergences that satisfy a ``mean-as-minimizer'' property
\cite{banerjee2005clustering}, while any divergence that satisfy the
data processing inequality is necessarily an $f$-divergence (or a
unique (one-to-one) mapping thereof) \cite{zakai1975generalization}.
As a consequence, any divergence that satisfies both of these
important properties simultaneously is necessarily proportional to the
KL divergence \cite[Corollary~6]{harremoes2007information}.
Additional properties of KL divergence are also discussed in
\cite{harremoes2007information}.

Finally, divergences inequalities have been studied extensively.  The
most celebrated example is the Pinsker inequality \cite{CsiszarIneq},
which expresses that KL divergence upper bounds the squared
total-variation distance.  More recently, the detailed studies of Reid
and Williamson \cite{reid2011information}, Harremo\"es and Vajda
\cite{harremoes2011pairs}, Sason and Verd\'u \cite{sason2016f}, and
Sason \cite{sason2018f} have extended this result to a broader set of
$f$-divergences inequalities.  Moreover, $f$-divergence inequalities
for non-probability measures appear in, e.g., by Stummer and Vajda
\cite{stummer2010divergences}.  In \cite{zhang2004statistical}, Zhang
demonstrated an important Bregman inequality in the context of
statistical learning, showing that the KL divergence upper bounds the
squared excess-risk associated with the 0-1 loss, and thus controls
this traditionally important performance measure.  Within this
context, our work can be viewed as extending such Bregman inequalities
and their analysis.

\section{Notation, Terminology and Definitions}
\label{basic}

Let $Y\in \{0,1\}$ be a Bernoulli random variable with parameter
$p=p_Y(1)$, which may be unknown.  A loss function $l(y,\yh)$
quantifies the discrepancy between a realization $Y=y$ and its
corresponding estimate $\yh$.  In this work we focus on probabilistic
estimates $\yh \defeq q \in [0,1]$ whereby $q$ is an estimate of $p$
rather than $y$ itself; as such, $q$ is a ``soft'' decision.

A loss function for such estimation takes the form
\begin{equation} 
l(y,q)=\kron\{y=0\}\, l_0(q)+\kron\{y=1\}\, l_1(q),
\label{eq:loss-form}
\end{equation}
with $\kron\{\cdot\}$ denoting the Kronecker (indicator) function,
where $l_k(q)$ is a loss function associated with the event $Y=k$, for
$k\in\{0,1\}$.  In turn, the corresponding expected loss is
\begin{equation} 
L(p,q) \defeq \E{l(Y,q)}=(1-p)\, l_0(q)+p\, l_1(q),
\label{eq:L-def}
\end{equation}
where we note that $L(p,q)$ depends only on $p$ and the estimate
$q$.   An example is the log-loss, for which
\begin{equation}
l_{\log}(y,q) \defeq y \log\frac1{q} + (1-y)\log\frac1{1-q}.
\label{eq:logloss}
\end{equation}

Loss functions with additional properties are of particular interest.
A loss function is \emph{proper} (or, equivalently,
\emph{Fisher-consistent} or \emph{unbiased}) if a minimizer of the
expected loss is the true underlying distribution of the random
variable we are to estimate; specifically,
\begin{equation} 
p\in\argmin_{q\in[0,1]}{L(p,q)}, \quad p\in[0,1].
\label{eq:proper}
\end{equation}
A \emph{strictly proper} loss function means that
$q=p$ is the unique minimizer, i.e.,
\begin{equation} 
p=\argmin_{q\in[0,1]}{L(p,q)}, \quad p\in[0,1].
\label{eq:strict-proper}
\end{equation}
A proper loss function is \emph{fair} if
\begin{align}
\label{fair}
l_0(0)=l_1(1)=0,
\end{align}
in which case there is no loss incurred for accurate
prediction.  Additionally, a proper loss function is
\emph{regular} if
\begin{equation}
\lim_{q\to0} q\, l_1(q) = \lim_{q\to1} (1-q)\, l_0(q) = 0.
\label{eq:regular}
\end{equation}
Intuitively, \eqref{eq:regular} ensures that making mistakes on events
that cannot happen do not incur a penalty.  

The minimum of the expected loss for proper loss functions, which we
denote using 
\begin{equation*}
G(p) \defeq L(p,p),
\end{equation*}
is referred to as the \emph{generalized entropy function}
\cite{gneiting2007strictly}, \emph{Bayes risk}
\cite{reid2010composite} or \emph{Bayesian envelope}
\cite{merhav1993universal}.  As an example, the Shannon
entropy associated with the log-loss \eqref{eq:logloss} is
\begin{equation} 
G_{\log}(p) \defeq p\log\frac1p + (1-p)\log\frac1{1-p}.  
\label{eq:shannon-ent}
\end{equation}

The \emph{regret} is defined as the difference between the expected
loss and its minimum, so for proper loss functions takes the form
\begin{equation} 
\Delta L(p,q)=L(p,q)-G(p).
\label{eq:regret}
\end{equation}
Savage \cite{savage1971elicitation} shows that a loss function
$l(y,q)$ is proper and regular if and only if $G(\cdot)$ is concave and
for every $p,q \in[0,1]$ we have that
\begin{equation}
L(p,q)=G(q)+(p-q)\,G'(q).
\label{eq:L-alt}
\end{equation}

This property allows us to draw an immediate connection between regret
and Bregman divergence.  In particular, let $f\colon\cS\mapsto \reals$ be
a continuously differentiable, strictly convex function over some
interval $\cS\subset \reals$.  Then its associated Bregman
divergence takes the form
\begin{equation} 
D_f(s\|t) \defeq f(s)-f(t)- (s-t)f'(t)
\label{eq:bregdiv-bin}
\end{equation}
for any $s,t \in \cS$.  We focus on closed intervals, in which case
the formal definition of $D_f(s\|t)$ at boundary points requires more
care; the details are summarized in
\appref{app:A}, following \cite{broniatowski2019some}.

In the special case $\cS = [0,1]$
using \eqref{eq:L-alt} in
\eqref{eq:regret} and comparing the result to \eqref{eq:bregdiv-bin} we
obtain
\begin{equation}
\Delta L(p,q) = D_{-G}(p,q),
\label{eq:regret-alt}
\end{equation}
i.e., the regret of a proper loss function is uniquely associated with
a Bregman divergence.  As an important example, associated with the
Shannon entropy \eqref{eq:shannon-ent} is the KL divergence
\begin{equation}
D_{\mathrm{KL}}(p\|q) \defeq p \log \frac{p}{q} + (1-p)\log\frac{1-p}{1-q}.
\end{equation}

Of particular interest are loss functions that are convex, i.e., $l$
such that $l(y,\cdot)$ is convex.  Such loss functions play a special
role in learning theory and optimization
\cite{buja2005loss,reid2010composite}.  For example,
suppose\footnote{The sequence notation $a^m=(a_1,\dots,a_m)$ is
  convenient in our exposition.}  $X^d$ and $Y$ is a set of $d$
explanatory variables (features) and a (target) variable,
respectively.  Then given a set of $n$ i.i.d.\ samples of $X^d$ and
$Y$, the empirical risk minimization (ERM) criterion seeks to minimize
\begin{equation*}
\frac{1}{n}\sum_{i=1}^n l(y_i,q_i),
\end{equation*}
where $q_i \defeq q_i(x^d_i)$ denotes a functional of the $i$\/th
sample of $X^d$.  This minimization is much easier to carry out when
the loss function $l$ is convex, particularly when $d$ is large.  In
addition, the minimum of the expected loss $L(p,\cdot)$ for a given
$p$ subject to constraints is typically much easier to characterize
and compute when $l$ is convex.  

Conveniently, convex proper loss functions $l(y,q)$ correspond to Bregman
divergences $D_{-G}$ such that $D_{-G}(p\|\cdot)$ is convex
\cite{reid2010composite}.  This family of divergences are of special
interest in many applications
\cite{bauschke2001joint,byrne2014iterative}, and have an important
role in our results, as will become apparent.

Accordingly, our development emphasizes the following class of analytically
convenient loss functions.
\begin{definition}
\label{def:admissible}
A loss function $l\colon\{0,1\}\times[0,1]\mapsto\reals$, which takes
the form \eqref{eq:loss-form}, is \emph{admissible} if it satisfies
the following three properties:
\begin{enumerate}[label=P1.\arabic*), wide=0pt, leftmargin=*, ref=P1.\arabic*]
\item $l(y,q)$ is strictly proper, fair, and regular, i.e., satisfies
  \eqref{eq:strict-proper}--\eqref{eq:regular}. \label{1}
\item $l(y,\cdot)$ is convex for each $y\in\{0,1\}$.  \label{2}
\item $l(y,\cdot)$ is in $\cC^3$ for each $y\in\{0,1\}$, i.e., 
  ${\p^k l(y,q)}/{\p q^k}$ exist and are
  continuous for $k=1,2,3$.  \label{3}
\end{enumerate}
\end{definition}
For convenience, we refer to loss functions that satisfy property
\ref{3} as \emph{smooth}.

As further terminology, for a proper, smooth loss function $l(y,q)$
with generalized entropy $G(p)$,
\begin{equation} 
w(p) \defeq - G''(p) 
\label{eq:wf-def}
\end{equation}
is referred to as its \emph{weight function}, which we note is
nonnegative.  As an example, that corresponding to the log-loss is
\begin{equation}
w_{\mathrm{KL}}(q)=\frac1{q(1-q)}.
\label{eq:wf-kl}
\end{equation}
Using \eqref{eq:wf-def}, we obtain, for example,
\begin{equation}
\frac{\p}{\p q} D_{-G}(p\|q) = (q-p)\, w(q),
\label{eq:D-grad-form}
\end{equation}
by differentiating \eqref{eq:L-alt}, which emphasizes the one-to-one
correspondence between $D_{-G}$ and $w$ for such loss functions; see
\appref{app:extra} for  additional properties and
characterizations.

Finally, representative examples of loss functions are provided in
\tabref{table:loss_functions}, along with their generalized entropies,
their associated Bregman divergences, and their weight functions.
\begin{table*}[tbp]
\caption{Examples of Commonly Used Binary Loss Functions
\label{table:loss_functions}}
\centering
\begin{tabular}{|c|M{2.8cm}|M{2.5cm}|M{5.5cm}|M{2.5cm}|N}
\hline

\xrowht{10pt}
Loss function
& $l(y,q)$ & $G(p)=L(p,p)$ & $D_{-G}(p\|q)$ & $w(p)$\\
\hline
\hline\xrowht{20pt}
\begin{tabular}{@{}c@{}} Quadratic loss\end{tabular}  
& \begin{tabular}{@{}c@{}} $y(1-q)^2+(1-y)q^2$\end{tabular} 
& \begin{tabular}{@{}c@{}} $p(1-p)$\end{tabular} 
&
\begin{tabular}{@{}c@{}} $D_{\mathrm{QL}}(p\|q)=(p-q)^2 $  \;\; \end{tabular} 
&
\begin{tabular}{@{}c@{}} $2$ \end{tabular}
&\\ 

\hline\xrowht{20pt}
\begin{tabular}{@{}c@{}} Logarithmic loss\end{tabular}  
& \begin{tabular}{@{}c@{}} $\begin{aligned} y&\log\ts
      \frac{1}{q}\\ &{}+(1\!-\!y)\log \ts\frac{1}{1-q}\\\ 
    \end{aligned}$\end{tabular} 
& \begin{tabular}{@{}c@{}} $\begin{aligned} p &\log\ts
      \frac{1}{p}\\&{}+(1-p)\log \ts \frac{1}{1-p}\\
    \end{aligned}$ \end{tabular} 
&
\begin{tabular}{@{}c@{}} $D_{\mathrm{KL}}(p\|q)=p \log \frac{p}{q} + (1-p)\log \frac{1-p}{1-q}$ \end{tabular} 
&
\begin{tabular}{@{}c@{}} $\frac{1}{p(1-p)}$ \end{tabular}
&\\ 

\hline\xrowht{25pt}
\begin{tabular}{@{}c@{}} Boosting loss\end{tabular}  
& \begin{tabular}{@{}c@{}} $\begin{aligned}
      2y&\ts\sqrt{\frac{1-q}{q}}\\&{}+
      2(1-y)\ts\sqrt{\frac{q}{1-q}}\\\ 
    \end{aligned}$\end{tabular} 
& \begin{tabular}{@{}c@{}} $4\sqrt{p(1-p)}$\end{tabular} 
& \begin{tabular}{@{}c@{}} $\begin{aligned} D_{\mathrm{BL}}(p\|q)
=2\biggl(p\ts\sqrt{\frac{1-q}{q}}+(1\!&-\!p)\ts\sqrt{\frac{q}{1-q}}
\biggr)\\
&{}-4\ts\sqrt{p(1-p)}\end{aligned}$\end{tabular}  
&
\begin{tabular}{@{}c@{}} $\frac{1}{(p(1-p))^{3/2}}$ \end{tabular}
&\\  \hline

\end{tabular}
\end{table*}

\section{Universality Properties of the Logarithmic Loss Function}
\label{main result}

Our main result is as follows, a proof of which is provided in
\appref{app:B}.   
\begin{theorem}
\label{main_theorem}
Given a loss function $l(y,q)$ satisfying \defref{def:admissible} with
corresponding generalized entropy function $G$, then for every $p,q
\in [0,1]$, 
\begin{subequations} 
\begin{equation}
D_{\mathrm{KL}}(p\|q)\ge \frac {1}{C(G)} D_{-G}(p\|q),
\label{main_ineq}
\end{equation}
where 
\begin{equation} 
C(G)>-\frac{1}{2}\, G''\biggl(\frac12\biggr)
\end{equation}
\end{subequations}
is a positive normalization
constant (that does not depend on $p$ or $q$).
\end{theorem}
Note that a further consequence of \thmref{main_theorem} expresses
that KL divergence is a ``dominating'' Bregman divergence in the sense
that given another Bregman divergence $\tilde{D}(p\|q)$ such that
[cf.\ \eqref{main_ineq}] 
\begin{equation*}
\tilde{D}(p\|q)\ge \frac {1}{\tilde{C}(G)} D_{-G}(p\|q)
\end{equation*}
holds for any Bregman divergence $D_{-G}$ for some $\tilde{C}(G)$,
then the theorem asserts that there exists $\tilde{C}_\mathrm{KL}$
such that
\begin{equation*}
D_\mathrm{KL}(p\|q) \ge \frac1{\tilde{C}_\mathrm{KL}}\,\tilde{D}(p\|q).
\end{equation*}
In essence, the dominating Bregman divergences form
  an equivalence class, of which KL divergence is a member.

We emphasize the necessity of scaling constants in
\thmref{main_theorem}. Indeed, the class of loss functions satisfying
\defref{def:admissible} is closed under (nonnegative) scaling, i.e.,
if $l(y,q)$ (with a corresponding $G$) satisfies
\defref{def:admissible}, then so does $\gamma l(y,q)$---with a
corresponding $\gamma G$---for any $\gamma>0$.  A typical approach to
placing loss functions on a common scale is to define a universal
scaling by setting, for instance,
\begin{equation*}
-\frac{1}{2}\, G''\biggl(\frac12\biggr)=1,
\end{equation*}
as appears, e.g., in \cite{buja2005loss, reid2011information}.
\thmref{main_theorem} avoids imposing such a normalization, and
instead absorbs such scaling into the constant $C(G)$ to obtain the
desired invariance.  As an example, for the quadratic loss
$G''(1/2)=-2$, so any $C(G)>1$ suffices in this case, whence
\begin{equation}
D_{\mathrm{KL}}(p\|q)\ge (p-q)^2.
\label{eq:quad-bin}
\end{equation}

The practical implications of \thmref{main_theorem} are quite
immediate.  Assume that the performance measure according to which a
learning algorithm is to be measured is unknown \textit{a priori} to
the application (as is the case, e.g., in the weather forecasting
example of \secref{intro}).  In such cases, minimizing the log-loss
provides an upper bound on any possible choice of measure that is
associated with an ``analytically convenient'' loss function.  As
such, the log-loss is a universal choice for classification problems
with respect to this class of measures.

More generally, as discussed in \secref{Previous Work}, designing
suitable loss functions is an active research field with many
applications.  Via \thmref{main_theorem}, one obtains universality
guarantees for any (current or future) loss function that is proper,
convex, and smooth.  We emphasize that this class of loss functions is
quite rich.  For instance, it is straightforward to verify that the
loss functions satisfying \defref{def:admissible} form a convex set:
any convex combination of such loss functions also satisfies
\defref{def:admissible}.

The local behavior of proper, convex, and smooth loss functions can be
derived from \thmref{main_theorem}.  In particular, we have the
following corollary.
\begin{corollary}
\label{local_analysis}
Given a loss function $l(y,q)$ satisfying \defref{def:admissible},
whose corresponding generalized entropy function is $G$, we have, for
every $p,p+\diff p \in [0,1]$ and some finite $C(G)>0$,
\begin{subequations} 
\begin{equation}
\frac{1}{C(G)} \, D_{-G}(p\|p+\diff p) \le \frac{\diff p^2}{2}\, J(p) +
o(\diff p^2),   
\end{equation}
where 
\begin{equation} 
J(p) \defeq \frac1{p(1-p)}
\label{eq:J-def}
\end{equation}
\end{subequations}
denotes the Fisher information of a Bernoulli distributed random
variable with parameter $p$.
\end{corollary}

\begin{IEEEproof}
With $p,p+\diff p \in [0,1]$, the Taylor series expansion of
the KL divergence around $p$ is
\begin{equation} 
D_{\mathrm{KL}}(p\|p+\diff p) =  \frac{\diff p^2}{2} J(p) + o(\diff p^2),
\label{coro_1}
\end{equation}
where $J(p)$ is as given in \eqref{eq:J-def}.  Substituting
\eqref{coro_1} into \eqref{main_ineq} yields the desired inequality.
\end{IEEEproof}

\corolref{local_analysis} establishes that when $q$ is sufficiently
close to $p$, the divergence associated with the set of smooth, proper
and convex binary loss functions is effectively upper bounded by
the Fisher information that locally characterizes KL divergence.  As
such, we conclude that the rate of convergence of any $D_{-G}(p\|q)$
to zero as $q\to p$ is upper bounded by the rate of
$D_{\mathrm{KL}}(p\|q)$.  This reveals that the price paid for the
universality of the log-loss is its slower rate of convergence.  
Such behavior will be demonstrated empirically in
\secref{illustrations}.

\section{Extended Bregman Divergence Inequalities}
\label{Bregman_Divergence_Inequalities}

To extend our result to arbitrary finite alphabets, we consider the
corresponding broader class of Bregman divergences.  In particular,
for a continuously differentiable, strictly convex function
$f\colon\cS\mapsto \reals$ be a over some convex set $\cS\subset
\reals^m$, its associated Bregman divergence takes the form
\begin{align}
D_f(s^m\|t^m) \defeq f(s^m)-f(t^m)-\ip{s^m-t^m}{\nabla f(t^m)}
\label{eq:bregdiv-m}
\end{align}
for any $s^m,t^m \in \cS$ when $\cS$ is open, where $\nabla f(t^m)$
is the gradient of $f$ at $t^m$.

We focus on the set $\cS=[0,1]^m$, and let $p^m,q^m\in\cS$.  We
emphasize that this is an extension beyond the unit simplex.  Let
\begin{equation}
H_f(p^m)  \defeq \nabla^2 f(p^m)
\label{eq:hessian}
\end{equation}
denote the $m\times m$ 
Hessian matrix of $f$.  For example, the divergence associated with
\begin{equation} 
f(p^m)=\sum_{i=1}^m p_i\log p_i
\label{eq:KL-f}
\end{equation}
is the \emph{generalized
  $\mathrm{KL}$ divergence}
\begin{equation}
\tilde{D}_{\mathrm{KL}}(p^m\|q^m) \defeq \sum_{i=1}^m p_i \log
\frac{p_i}{q_i} -\sum_{i=1}^m p_i +\sum_{i=1}^m q_i,
\label{eq:KL-gen}
\end{equation}
the corresponding Hessian for which is
\begin{equation*}
H_{\mathrm{KL}}(p^m) \defeq \nabla^2 \left(\sum_{i=1}^m p_i \log p_i\right),
\end{equation*}
which we note
is a diagonal matrix whose $i$\/th diagonal element is $1/p_i$.  In
the special case wherein $p^m$ and $q^m$ are probability measures (i.e.,
restricted to the unit simplex), we have
\begin{equation*}
\tilde{D}_{\mathrm{KL}}(p^m\|q^m) = D_{\mathrm{KL}}(p^m\|q^m)
\defeq \sum_{i=1}^m
p_i \log \frac{p_i}{q_i},
\end{equation*}
which generalizes the definition in
\tabref{table:loss_functions}.  

We focus on the following class of Bregman divergences.
\begin{definition}
\label{def:admissible-m}
For some integer $K$, a Bregman divergence generator $f\colon [0,1]^m
\mapsto \reals$ is $K$-admissible if it satisfies the following
properties:
\begin{enumerate}[label=P2.\arabic*), wide=0pt, leftmargin=*, ref=P2.\arabic*]
\item $f$ is a strictly convex function
  that is well-defined on its boundaries, in the sense of 
  generalizing the requirements of 
  \appref{app:A}.\label{I} 
\item $f\in\cC^K$, i.e.,  ${\p^k f(p^m)}/{\p p_{1} \cdots \p p_k}$
  exist and are continuous for $k = 1,\dots, K$.\label{II}
\end{enumerate}
\end{definition}

Our first generalization is the following theorem, whose proof is
provided in \appref{app:C}.
\begin{theorem}
\label{divergence_theorem_general}
Given a positive integer $m$, let $f\colon [0,1]^m \mapsto \reals$
satisfy \defref{def:admissible-m} for $K=2$, and let
$D_f(p^m\|q^m)$ and $H_f(p^m)$ denote the associated Bregman
divergence and Hessian matrix, respectively.  If there exists a
(finite) positive constant $C(f)$ such that\footnote{We use $A \succ
  0$ to denote that a matrix $A$ is positive definite.}
\begin{subequations} 
\begin{equation} 
C(f)\, H_{\mathrm{KL}}(p^m)-H_f(p^m) \succ 0, 
\quad \text{all $p^m \in  [0,1]^m$},
\label{pd_condition}
\end{equation}
then for every $p^m, q^m \in [0,1]^m$, 
\begin{equation}
\tilde{D}_{\mathrm{KL}}(p^m\|q^m) \ge \frac{1}{C(f)} \, D_f(p^m\|q^m).
\label{main_ineq_div_general}
\end{equation}
\end{subequations}
\end{theorem}

We emphasize that, in contrast to \thmref{main_theorem}, the
inequality \eqref{main_ineq_div_general} applies to any Bregman
divergence satisfying \defref{def:admissible-m}, and in particular
does not require $D_f(p^m\|\cdot)$ to be convex for any
$p^m\in[0,1]^m$.  However, at the same time, we stress that
\thmref{divergence_theorem_general} is restricted to the class of
divergences satisfying \eqref{pd_condition}.

As an example application, when\footnote{Here, and elsewhere as
  needed, we construe a sequence $a^m$ as a column vector.}
\begin{equation}
f(p^m)=(p^m)^\T Q\, p^m,
\label{eq:f-mahalanobis}
\end{equation}
with positive definite matrix parameter $Q$,
the corresponding the Bregman divergence
\begin{equation*} 
D_f=\frac{1}{2}(p^m-q^m)^\T Q \, (p^m-q^m)
\end{equation*}
is the well-known Mahalanobis distance, and and the associated Hessian
is
\begin{equation*} 
H_f(p^m)=Q.
\end{equation*}
For this divergence we have the following corollary, whose proof is
provided in \appref{app:D}.
\begin{corollary}
\label{corol:mahalanobis}
If $D_f$ is a Mahalanobis distance, whereby $f$ takes the form
\eqref{eq:f-mahalanobis} with $Q\succ0$, 
then \eqref{main_ineq_div_general} holds for 
\begin{equation} 
C(Q)>\lambda_{\max}(Q),
\label{eq:mahalanobis-cond}
\end{equation}
where $\lambda_{\max}(Q)$ is the largest eigenvalue of $Q$.
\end{corollary}

Our second generalization of \thmref{main_theorem} focuses on the
class of \emph{separable} Bregman divergences, a member of which
takes the form
\begin{subequations} 
\begin{equation} 
D_g(p^m\|q^m) \defeq \sum_{i=1}^{m} d_g(p_i\|q_i)
\label{eq:breg-sep}
\end{equation}
with
\begin{equation}
d_g(p_i\|q_i) \defeq g(p_i)-g(q_i)-(p_i-q_i)\,g'(q_i),
\label{eq:dg-def}
\end{equation}
\end{subequations}
for $p^m,q^m\in(0,1)^m$, where $g\colon[0,1]\mapsto \reals$ denote a
continuously differentiable, strictly convex function with additional
constraints discussed analogous to those discussed in \appref{app:A},
and via which \eqref{eq:breg-sep} is extended to $p^m,q^m\in[0,1]^m$.

Such divergences hold a special role in divergence analysis, as
discussed in, e.g., \cite{harremoes2007information,jiao2014information}.
Note that in this case, the Bregman generator function takes the form
\begin{equation}
f_\mathrm{KL}(p^m) = \sum_{i=1}^m g(p_i),
\label{eq:f-KL-def}
\end{equation}
via which we obtain the Hessian as
\begin{equation*}
H_f(p^m) = \kron\{i=j\}\, g''(p_i).
\end{equation*}
As an example,
\begin{equation} 
g(p) = p\log p
\label{eq:KL-g}
\end{equation}
matches \eqref{eq:KL-f}, and when used in \eqref{eq:dg-def} yields
\begin{equation}
\tilde{d}_\mathrm{KL}(p_i\|q_i) \defeq p_i\log\frac{p_i}{q_i} - p_i + q_i,
\label{eq:dKL-def}
\end{equation}
so that \eqref{eq:breg-sep} specializes to the generalized KL
divergence \eqref{eq:KL-gen}.

Our main result is the following theorem, a proof of which is
provided in \appref{app:E}.
\begin{theorem}
\label{divergence_theorem}
Given a positive integer $m$, let $D_g(p^m\|q^m)$ be a separable
Bregman divergence satisfying \defref{def:admissible-m} for $K=3$ and
for which $D_g(p^m\|\cdot)$ is convex for every $p^m\in[0,1]^m$.  Then
for every $p^m, q^m \in [0,1]^m$,
\begin{subequations} 
\begin{equation}
\tilde{D}_{\mathrm{KL}}(p^m\|q^m) \ge \frac{1}{C(g)} D_g(p^m\|q^m)
\label{main_ineq_div}
\end{equation}
when $C(G)$ satisfies
\begin{equation}
C(g)>g''(1).
\label{eq:Cg-cond}
\end{equation}
\end{subequations}
\end{theorem}
We remark that when $g''(1)$ is unbounded,
\corolref{divergence_theorem} does not yield a useful bound.  By
contrast, \thmref{main_theorem} is guaranteed to produce a bound,
since $G''(1/2)$ is always finite.

It is important to emphasize that while \thmref{main_theorem}
restricts attention to divergences defined both over binary alphabets
and only on the unit simplex---i.e., in the notation of this
section,
\begin{equation*}
m=2,\qquad p_1 = p,\qquad  p_2 = 1-p,\qquad p\in[0,1],
\end{equation*}
by contrast the divergences in \thmref{divergence_theorem} are defined
for any positive integer $m$ and, in addition, over the entire
hypercube $p^m,q^m \in [0,1]^m$.  As such, we emphasize that
\thmref{main_theorem} is not a special case of
\thmref{divergence_theorem}.  In particular, because
\eqref{main_ineq} must hold for a domain that extends beyond
the unit simplex, the smallest $C(g)$ for which it is
satisfied when $m=2$ must generally be bigger than the smallest
$C(G)$ for which \eqref{main_ineq}
holds.\footnote{That said, if desired, via similar analysis, together
  with the use of Lagrange multipliers, one can obtain a version of
  \thmref{divergence_theorem} restricted to the unit simplex, for
  which smaller constants will generally be obtained.}

As a simple application of \thmref{divergence_theorem}, choosing
$g(p)=p^2$ generates the quadratic divergence
\begin{equation*}
D_g(p\|q) = \sum_{i=1}^m (p_i-q_i)^2,
\end{equation*}
which is a special case of the Mahalanobis distance.  In this case,
since $g''(1)=2$, \thmref{divergence_theorem} requires $C(g)>2$, 
yielding
\begin{equation} 
\tilde{D}_{\mathrm{KL}}(p^m\|q^m) \ge \frac{1}{2} \sum_{i=1}^m (p_i-q_i)^2.
\label{mse}
\end{equation}
Consistent with the preceding discussion, $\inf\{C(g)\colon
C(g)>2\}=2$ corresponding to \eqref{mse}, is larger than the
corresponding $\inf\{C(G)\colon C(G)>1\}=1$ in the bound \eqref{eq:quad-bin}.

Additionally, it is worth noting that \eqref{mse} resembles the
well-known Pinsker inequality \cite{cover2012elements}, viz.,
\begin{equation} 
    D_{\mathrm{KL}}(p^m\|q^m) \ge \frac{1}{2} D^2_{\mathrm{TV}}(p^m\|q^m),
\label{pinsker}
\end{equation}
where
\begin{equation} 
D_{\mathrm{TV}}(p^m\|q^m) \defeq \sum_{i=1}^m |p_i-q_i|
\label{eq:tv-def}
\end{equation}
is the total-variation distance (or Csisz\'ar divergence
\cite{cover2012elements}), which is not a Bregman divergence, but
rather an $f$-divergence.  It is straightforward to verify that
\eqref{pinsker} is tighter than \eqref{mse} when $p^m$ and $q^m$ are
restricted to the unit simplex.  Nevertheless, this simple example
serves to illustrate that \thmref{divergence_theorem_general} and
\thmref{divergence_theorem} may be viewed as Bregman divergences
extensions to some well-known $f$-divergence results, as discussed in
\secref{Previous Work}.

\section{Numerical Analysis and Experiments}
\label{illustrations}

To complement the results of \secref{Bregman_Divergence_Inequalities},
we use numerical analysis to examine the dependence of
\begin{equation*}
\tilde{D}_{\mathrm{KL}}(p^m\|q^m) -
\frac {1}{C(f)} D_{f}(p^m\|q^m)
\end{equation*}
on $p^m$ and $q^m$, for some choices of $f$ such that
\thmref{divergence_theorem_general} applies, and with $C(f)$ chosen
according to \eqref{pd_condition}.

To begin, we consider a random variable
$Y\in\cY$ with $|\cY|=m$, and restrict $q^m$ to lie in a subset $\cS$
of the unit simplex $\cP^\cY$.  For a given $p^m\in\cP^\cY$, with
\begin{equation} 
q^m_f(\cS) \defeq \argmin_{\{q^m\in\cS\subset\cP^\cY\}}
D_f(p^m\|q^m) \label{eq:qmf-def}
\end{equation}
and, in turn,
\begin{equation} 
q^m_\mathrm{KL}(\cS) \defeq q^m_{f_\mathrm{KL}}(\cS),
\label{eq:qmKL-def}
\end{equation}
where $f_\mathrm{KL}$ is as given in \eqref{eq:f-KL-def}, 
the minimum KL divergence upper bounds the minimum of any 
Bregman divergence according to
\begin{align} 
D_{\mathrm{KL}}(p^m\|q_{\mathrm{KL}}^m(\cS)) &\ge
\frac{1}{C(f)}D_f(p^m\|q_{\mathrm{KL}}^m(\cS)) \notag\\
&\qquad\qquad{}\ge 
\frac{1}{C(f)}D_f(p^m\|q_f^m(\cS)),
\label{examples1}
\end{align}
where to obtain the first inequality we have used
\thmref{divergence_theorem_general} with $C(f)$ satisfying
\eqref{pd_condition}, and to obtain the second inequality we have used
\eqref{eq:qmf-def}. 

\begin{figure}[tbp]
\centering
\includegraphics[width =0.42\textwidth,bb= 80 190 510 570,clip]{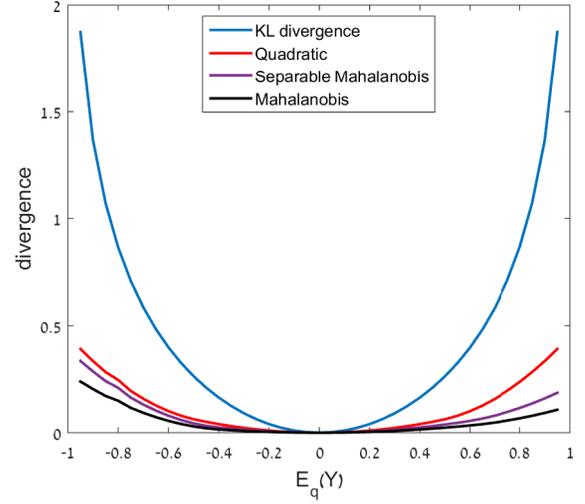}
\caption{Divergence bound behavior under mean constraint. Depicted is
  the minimum of $D_f(p^3\|q^3)/C(f)$ with respect to $q^3\in\cP^\cY$
  with $\Ed{q^3}{Y}$ fixed, for $Y\in\cY=\{-1,0,1\}$, $p^3=\left(
  1/4,1/2,1/4\right)$, and different choices for $f$, as described in
  text.
\label{fixed_stat_experiment}}
\end{figure}

In the first experiment, we set
\begin{equation*}
\cS = \bigl\{ q^m\in\cP^\cY\colon \Ed{q^m}{h(Y)}=\mu \bigr\},
\end{equation*}
for some $h\colon\cY\mapsto \reals$ and $\mu$, i.e., we constrain
$q^m$ to lie in a hyperplane restricted to the unit simplex $\cP^\cY$.
More specifically, we choose $\cY=\{-1,0,1\}$, $h(y)=y$, and
$p^3=(1/4,1/2,1/4)$ to illustrate our results.  The results of our
experiment, which compares the minima in \eqref{examples1} as a
function of $\mu$, are depicted in \figref{fixed_stat_experiment}.
The top (blue) curve is (with a minor abuse of notation)
$D_{\mathrm{KL}}(p^3\|q^3_\mathrm{KL}(\mu))$, and the progressively
lower (red, purple, and black) curves are (with a similar abuse of
notation)  $D_f(p^3\|q^3_f(\mu))/C(f)$ with
$f$ corresponding to the quadratic divergence, the separable
Mahalanobis distance with parameters $Q_\mathrm{s}$, and the
nonseparable Mahalanobis distance with parameters $Q_\mathrm{ns}$,
respectively. The specific values of these parameters are
\begin{equation}
Q_\mathrm{s} = \begin{bmatrix} 3 & 0 & 0 \\ 0 & 2 & 0 \\ 0 & 0 & 1 
\end{bmatrix}\quad\text{and}\quad
Q_\mathrm{ns} = \begin{bmatrix} 3 & 1/2 & 1/2 \\ 1/2 & 2 & 1/2 \\ 1/2
  & 1/2 & 1  \end{bmatrix}.
\label{eq:mahalanobis-params}
\end{equation}

Note that since $\Ed{p^3}{Y}=0$ for our choice of $p^3$, all the
minimum divergences are zero at $\mu=\Ed{q^3}{Y}=0$, and thus
$\smash{q^3_f(0)=p^3}$ for all Bregman generators $f$.  However, when $\mu\ne0$,
the optimizing $q^3_f(\mu)$ must differ from $p^3$, and
\figref{fixed_stat_experiment} quantifies these differences as a
function of the bias $\mu$.  Consistent with the analysis of
\secref{Bregman_Divergence_Inequalities}, KL divergence upper bounds
normalized measures of all these differences.

\begin{figure*}[tbp]
\centering
\includegraphics[width =0.7\textwidth,bb= 40 110 690
  510,clip]{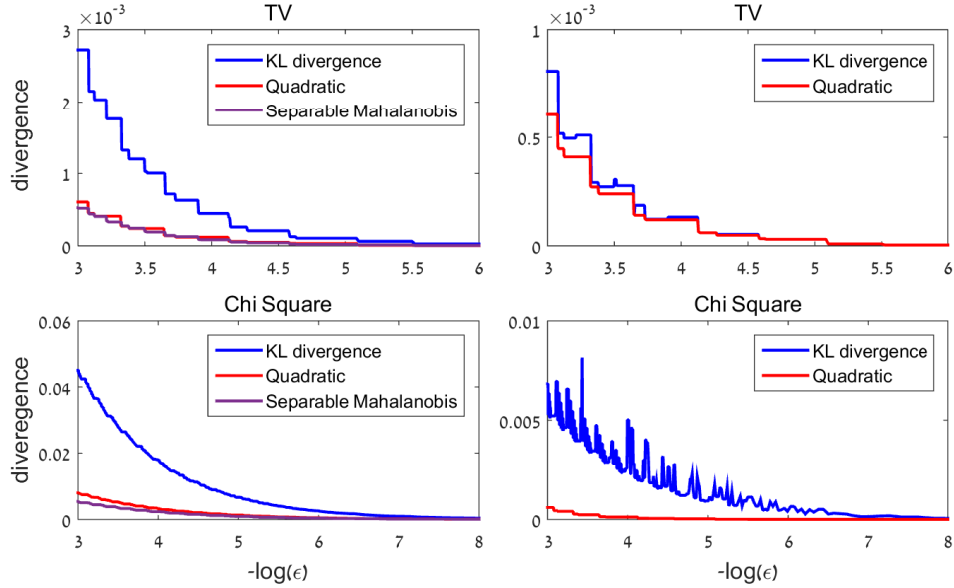} 
\caption{Divergence bound behavior under a divergence constraint.  On
  the left is depicted the minimum of $D_f(p^3\|q^3)/C(f)$ over $q^3$
  for $p^3=(1/4,1/2,1/4)$ subject to $\smash{D(p^3\|q^3)\ge \epsilon}$.  In
  the plots on the right, the minimizing $\smash{q^3}$ is replaced with that
  minimizing KL divergence.  The upper and lower plots correspond to
  $D$ being total-variation and chi-square divergences, respectively.
  The different choices for $f$ are as in
  \protect\figref{fixed_stat_experiment}.
\label{epsilon_experiment}}
\end{figure*}

In the second experiment we show that the bounds \eqref{examples1}
hold for a broader range of problems.  To model a
statistical, computational, or even algorithmic constraint that
prevents $q^m$ from converging to some given $p^m\in\cP^\cY$, we
impose that $q^m\in\cS$ where
\begin{equation}
\cS = \{ q^m\in \cP^\cY \colon D(p^m\|q^m) \ge \epsilon \}
\label{eq:cS-div}
\end{equation}
for some $D$ and $\epsilon>0$.  In
\figref{epsilon_experiment}, we compare the terms in \eqref{examples1}
for different choices for $f$, and 
two different (non-Bregman) examples of $D$ in \eqref{eq:cS-div}.  In
particular, the upper plots corresponds to choosing for $D$ in
\eqref{eq:cS-div} the total-variation distance $D_\mathrm{TV}$ as
defined in \eqref{eq:tv-def}. For constrast, the lower plots
corresponds choosing for $D$ in \eqref{eq:cS-div} the (Neyman)
chi-square divergence, i.e.,
\begin{equation}
D_{\chi^2}(p^m\|q^m)=\sum_{i=1}^m \frac{(p_i-q_i)^2}{q_i}.
\end{equation}

The plots on the left compare
$D_\mathrm{KL}(p^m\|q_\mathrm{KL}^m(\cS)$ with \eqref{eq:qmKL-def} to
$D_f(p^m\|q_f^m(\cS)/C(f)$ with \eqref{eq:qmf-def}, for $f$
corresponding to the quadratic and separable Mahalanobis distances
(where the latter has parameters $Q_\mathrm{s}$ as specified in
\eqref{eq:mahalanobis-params}).  Consistent with \eqref{examples1},
$D_\mathrm{KL}(p^m\|q_\mathrm{KL}^m(\cS))$ upper bounds
$D_f(p^m\|q_f^m(\cS))$ for both the quadratic divergence and the
separable Mahalanobis distance.  Moreover, we see that larger values
of $\epsilon$ result in a greater bias, as we would expect.

The plots on the right compare
$D_{\mathrm{KL}}(p^m\|q_\mathrm{KL}^m(\cS))$ with \eqref{eq:qmKL-def}
to the middle term in \eqref{examples1}, i.e.,
$D_f(p^m\|q_\mathrm{KL}^m(\cS))/C(f)$, for $f$ corresponding to the
quadratic distance.  The results demonstrate that
$q_{\mathrm{KL}}^m(\cS)$ can, indeed, be an effective approximation to
$q_f^m(\cS)$ with respect to minimizing $D_f(p^m\|\cdot)$.

In the third experiment we demonstrate the application of our bounds
to weather forecasting as discussed in \secref{intro}.  Recall that
weather forecasters typically assign probabilistic estimates to future
meteorological events.  The estimates are designed to minimize a
performance measure, according to which the weather forecaster is
evaluated.  However, weather estimates serve a wide audience, within
which different recipients may be interested in different and often
conflicting measures.  For example, by minimizing the quadratic loss,
a forecaster may reasonably assign zero probability of occurrence to
very rare events, but this would result in an unbounded logarithmic
loss.

To demonstrate the value of using log-loss minimization to control a
large set of commonly used performance measures, we analyze weather
data collected by the Australian Bureau of Meteorology \cite{anms17}.
This publicly available dataset contains the observed weather and its
corresponding forecasts in multiple weather stations in Australia.  In
our experiment we focus on the predicted chances of rain (where a
rainfall is defined as over 2mm of rain) compared with the true event
of rain.  Our dataset contains $n=33\,134$ pairs
$\{(x_1,y_1),\dots,(x_n,y_n)\}$ of forecasts and corresponding weather
observations that were collected during the period Apr.~28--30, 2016.
For reference, in this period, a fraction
\begin{equation*}
\frac1n \sum_{i=1}^n y_i = 0.09
\end{equation*}
of the observations correspond to an event of rain.  We evaluate the
accuracy of the Australian weather forecasts by the three commonly
used proper loss measures: logarithmic, quadratic, and 0-1 losses,
with the latter defined via
\begin{equation*}  
l(y,q) = y\,\kron\{q<t\} + (1-y)\, \kron\{q \ge t\},
\end{equation*}
and where we choose as its parameter $t=0.35$, following the Bureau's
guidelines.  The first row of \tabref{table:weather} summarizes our
results.

\begin{table}[tbp]
\caption{Weather Forecast Experiment \label{table:weather}}
\renewcommand{\baselinestretch}{1}\footnotesize
\centering
\begin{tabular}{|c|M{1.5cm}|M{1.5cm}|M{1.5cm}|N}  

\hline\xrowht{15pt}

Weather Forecaster
& 0-1 loss
& Quadratic loss 
& Logarithmic loss &\\

\hline
\hline\xrowht{15pt}

\begin{tabular}{@{}c@{}} Australian \\ Forecaster\end{tabular}  
& $0.0898$
& $0.0676$ 
& $\infty$ &\\\hline\xrowht{15pt}

\begin{tabular}{@{}c@{}} Modified \\ Forecaster\end{tabular}  
& $0.0901$
& $0.0675$ 
& $0.234$ &\\ \hline

\end{tabular}
\end{table}

Note that the unbounded logarithmic loss is a consequence of the fact
that there are several instances in which the forecaster predicted
zero chance of rain but it ultimately rained.  In correspondence with
them, Australia's National Meteorological Service confirmed that their
forecasts are typically internally evaluated by both a quadratic loss
and a 0-1 loss with parameter $t=0.35$.  In addition, they perform
more sophisticated evaluation analysis which is not in the scope of
this work.

Next, we consider a method for revising the existing forecasts based
on our log-loss universality results.  Since the available forecasts
are generated by a prediction algorithm whose features unavailable to
us, our revised forecasts can only be based on the existing
forecasts.  Accordingly, we make use of a simple logistic regression
in which the target is the observed data and the single feature
is the corresponding original forecast.  Specifically, given an
original weather forecast of $x\in[0,1]$, we generate the following
updated weather forecast according to
\begin{equation}
q_{\beta_0,\beta_1}(x)=\frac{1}{1+\e^{-\beta_0-\beta_1 x}},
\label{eq:logistic}
\end{equation}
where the regression parameters $\beta_0$ and $\beta_1$ are 
fit to training data
$\{(\tilde{x}_1,\tilde{y}_1),\dots,
(\tilde{x}_{\tilde{n}},\tilde{y}_{\tilde{n}})\}$ 
according to 
\begin{equation*}
\argmin_{\beta_0,\beta_1} \frac1{\tilde{n}} \sum_{i=1}^{\tilde{n}}
  l_{\log}\bigl(\tilde{y}_i,q_{\beta_0,\beta_1}(\tilde{x}_i)\bigr).
\end{equation*}
To avoid over-fitting, the training data was from Jan.~2016, and thus
different from the test data .  The accuracy of the resulting updated
forecasts are presented in the second row of \tabref{table:weather}.

Note that the updated forecasts now incur a bounded log-loss, and that
this robustness is achieved without significantly affecting accuracy
with respect to the other loss functions.  Evidently, even such simple
post-processing improves log-loss performance while controlling a
large set of alternative measures, consistent with the results of
\thmref{main_theorem} (and of those in \cite{zhang2004statistical} for
the 0-1 loss).

\section{Example Applications}
\label{application}

The Bernoulli log-likelihood loss function is widely used in a variety
of scientific fields.  Several key examples, in addition to those
discussed above, include logistic regression in statistical analysis
\cite{friedman2001elements}, the info-max criterion in machine
learning \cite{linsker1988self}, independent component analysis in
signal processing
\cite{painsky2016generalized},\cite{painsky2018linear}, splitting
criteria in classification trees \cite{breiman1984classification}, DNA
sequence alignment \cite{keith2002sequence}, and many others.  In this
section we demonstrate the potential applicability of our universality
results in the context of three key examples.    

\subsection{Universal Clustering with Bregman Divergences}

Data clustering is an unsupervised learning procedure that has been
extensively studied across a variety of disciplines over many decades.
Most clustering methods assign each data sample to one of a
pre-specified number of partitions, with each partition defined by a
cluster \emph{representative}, and where the quality of clustering is
measured by the proximity of samples to their assigned cluster
representatives, as measured by a pre-defined distance function.

Several popular algorithms for data clustering have been developed
over the years.  This includes the well-known $k$-means
algorithm \cite{macqueen1967some} which minimizes the quadratic distance.
Another widely used example is the Linde-Buzo-Gray (LBG) algorithm
\cite{linde1980algorithm,buzo1980speech} based on the Itakura-Saito
distance \cite{itakura1968analysis}.  More recently, Dhillion
\textit{et al.}\ \cite{dhillon2003divisive} proposed an
information-theoretic approach to clustering probability
distributions based on KL divergence.  

All of these clustering methods are based on an
Expectation-Maximization (EM) framework for minimizing the aggregate
distance, and share the same optimality property: the centroid
(representative) of each cluster is the mean of the data points that
are assigned to it.  Moreover, all of these algorithms use a Bregman
divergence as their measure of distance, as do some promising emerging
methods.  For example, a new class of clustering methods has been
shown to offer significant improvement in various domains by
utilizing so-called total Bregman divergence, a rotation-invariant
version of classical Bregman divergence \cite{liu2011total,
  liu2010total, vemuri2010total, liu2012shape, nock2015conformal}.

The connection between clustering and the Bregman divergence is
developed in Banerjee \textit{et
  al.}\ \cite{banerjee2005clustering}.  In particular, a key result is
that a random variable $X$ satisfies 
\begin{equation} 
\E{X}=\argmin_z \E{D_f(X\|z)}
\end{equation}
if and only if $D_f$ is a Bregman divergence.  It follows that any
clustering algorithm that satisfies the ``mean-as-minimizer'' property
centroid property minimizes a Bregman divergence, and thus we need
look no further than among the Bregman divergences in selecting a
candidate distance measure for EM-based data clustering.  

Even with this restriction, it is frequently not clear how to choose
an appropriate Bregman divergence for a given clustering task.
Banerjee \textit{et al.}\ \cite{banerjee2005clustering} show that
there is a unique correspondence between exponential families and
Bregman divergences.  As such, if the data are from an exponential
family, with different parameters for different clusters, then the
natural distance for clustering is the corresponding Bregman
divergence.  As an example, for Gaussian distributions with differing
means, the quadratic distance used by $k$-means is the natural
distance.  However, in practice, information about the generative
model for the data is rarely known.

As an alternative, our results suggest a ``universal'' approach to
clustering that provides performance guarantees with respect to any
Bregman divergence that might turn out to be relevant.  Specifically,
suppose we are given samples $x^n$ to be partitioned into $k$ clusters
with corresponding representatives $\mu^k=(\mu_1, \dots, \mu_k)$.
Then the optimum solution for measure $D_f$ is
\begin{equation*}
\mu_f^k \defeq \argmin_{\mu^k} \sum_{j=1}^k \sum_{i\in\cI^f_j(\mu^k)}
D_f(x_i\|\mu_j),
\end{equation*}
where\footnote{When a sample is equidistant to multiple
  representatives, we pick one arbitrarily.}
\begin{align*} 
&\cI^f_j(\mu^k) \notag\\
&\ \defeq \bigl\{ i\in\{1,\dots,n\} \colon 
D_f(x_i\|\mu_j)< D_f(x_i\|\mu_{j'}),\ \text{all $j'\ne j$}
 \bigr\}.
\end{align*}
Similarly, for measure $\tilde{D}_\mathrm{KL}$, we use the (slightly
simpler) notation $\cI^\mathrm{KL}_j(\mu^k) =
\cI^{f_\mathrm{ML}}_j(\mu^k)$, and $\mu_\mathrm{KL}^k
\defeq\mu_{f_\mathrm{KL}}^k$.

Using \thmref{divergence_theorem_general} (for $f$ and $C(f)$
satisfying the conditions of the theorem), we can then bound
performance with respect to $D_f$ according to [cf.\ \eqref{examples1}]
\begin{align}
&\sum_{j=1}^k \sum_{i\in\cI^\mathrm{KL}_j(\mu^k_\mathrm{KL})}
\tilde{D}_{\mathrm{KL}}(x_i\|\mu^\mathrm{KL}_j)\notag\\ 
&\qquad\qquad\ge \frac{1}{C(f)} \sum_{j=1}^k
\sum_{i\in\cI^\mathrm{KL}_j(\mu^k_\mathrm{KL})} D_f(x_i\|\mu^\mathrm{KL}_j) \notag\\ 
&\qquad\qquad\qquad\qquad\ge \frac{1}{C(f)} \sum_{j=1}^k \sum_{i\in\cI^f_j(\mu_f^k)} D_f(x_i\|\mu^f_j), 
\label{clustering_theorem}
\end{align} 
where $\mu_f^k = (\mu_1^f,\dots,\mu_k^f)$.

Via \eqref{clustering_theorem}, we conclude that by applying a
clustering algorithm that minimizes KL divergence, we provide
performance guarantees for any (reasonable) choice of clustering
method.  As such, our analysis provides further justification for the
popularity of the KL divergence in distributional clustering
\cite{pereira1993distributional} and specifically in the context of
natural language processing and text classification
\cite{baker1998distributional,clark2001unsupervised,bekkerman2001feature}.

\subsection{The Universality of the Information Bottleneck}

The information bottleneck \cite{tishby1999information} is a
conceptual machine learning framework for extracting an informative
but compact representation of an explanatory variable\footnote{Note
  that $X$ can equivalently represent a collection of variables.} $X$
with respect to inferences about a target $Y$, generalizing the notion
of a minimal sufficient statistic from classical parametric
statistics.  Given the joint distribution $p_{X,Y}$, the method
selects the compressed representation for $X$ that preserves the
maximum amount of information about $Y$.  As such, $Y$ effectively
regulates the compression of $X$, so as to maintain a level of
explanatory relevance with respect to $Y$.  Specifically, with $T$
denoting the compressed representation, the information bottleneck
problem is
\begin{equation} 
\max_{p_{T|X}} I(T;Y) \quad \text{subject to} \quad
I(X;T) \le \bar{I},
\label{eq:ib}
\end{equation}
where $T\markov X\markov Y$ form a Markov chain, and thus the
minimization is over all possible (generally randomized) mappings of
$X$ to $T$.  Here, $\bar{I}$ is a constant parameter that sets the level
of compression to be attained.  As
$\bar{I}$ is varied, the tradeoff between $I(X;T)$ (corresponding to
the representation complexity) and $I(T;Y)$ (corresponding to the
predictive power) is a continuous, concave function.

Information bottleneck analysis is a powerful tool in a variety of
machine learning domains and related areas; see, e.g.,
\cite{slonim2000document,friedman2001multivariate,sinkkonen2002clustering,slonim2005information,hecht2009speaker}.
It is also applicable in a variety of other fields, including
neuroscience \cite{schneidman2001analyzing} and optimal control
\cite{tishby2011information}.  Recently, there have been
demonstrations of its ability to analyze the performance of deep
neural networks \cite{tishby2015deep, shwartz2017opening,
  goldfeld2018estimating}. 

It is useful to recognize that the information bottleneck problem
\eqref{eq:ib} is an instance of a remote-source rate-distortion
problem \cite{cover2012elements}.  In particular, let $Y$ be a remote
source that is unavailable to the encoder, and let $X$ be a random
variable that is dependent of $Y$ through a (known) mapping $p_{X|Y}$,
which is available to the encoder.  The remote source coding problem
is to achieve the highest possible compression rate for $X$ given a
prescribed maximum tolerable reconstruction error of $Y$ from the
compressed representation $T$.  In this setting, the reconstruction
error is measured by a predefined distortion (loss) function, where
the choice of log-loss leads to the standard information
bottleneck problem \cite{shkel2017single}.

While the choice of log-loss is typically justified by several
properties of KL divergence \cite{harremoes2007information}, the
results of this paper can be applied to show that its use provides
valuable universality guarantees for the remote source coding problem.

To develop this view, first note that 
\begin{equation*} 
I(T;Y) = I(X;Y) -
\Ed{p_{X,T}}{D_\mathrm{KL}(p_{Y|X}(\cdot|X)\|p_{Y|T}(\cdot|T))},
\end{equation*}
which follows from straightforward algebra.  In this form, we
recognize $p_{Y|X}$ as the full predictive model and $p_{Y|T}$ as the
compressed predictive one.  Since $p_{X,Y}$ is given, we maximize
$I(T;Y)$ (as \eqref{eq:ib} dictates) by minimizing
$\Ed{p_{X,T}}{D_\mathrm{KL}(p_{Y|X}(\cdot|X)\|p_{Y|T}(\cdot|T))}$.  In
the more general souce coding problem, we instead seek to minimize
\begin{equation}
\Ed{p_{X,T}}{D_f(p_{Y|X}(\cdot|X)\|p_{Y|T}(\cdot|T))},
\label{eq:EDf}
\end{equation}
with $f$ chosen as desired.

When the appropriate choice of $f$ is not clear, via
\thmref{divergence_theorem_general} we have 
\begin{align}
&\Ed{p_{X,T}}{D_{f}(p_{Y|X}(\cdot|X)\|p_{Y|T}(\cdot|T))} \notag\\
&\quad\qquad\qquad\le C(f)\,
\Ed{p_{X,T}}{D_{\mathrm{KL}}(p_{Y|X}(\cdot|X)\|p_{Y|T}(\cdot|T))}  
\label{eq:ib-bound}
\end{align} 
for any Bregman divergence $D_f$ that satisfies
\defref{def:admissible-m} and \eqref{pd_condition} for some $C(f)>0$.
Therefore, by minimizing
$\Ed{p_{X,T}}{D_{\mathrm{KL}}(p_{Y|X}(\cdot|X)\|p_{Y|T}(\cdot|T))}$ we
effectively minimize \eqref{eq:EDf} for any divergence that might
reasonably be of interest.  

Finally, it is worth noting that in classification problems,
separable divergence measures are popular.  In this case, then, via
\thmref{divergence_theorem} we obtain a universality bound of the form
\eqref{eq:ib-bound} for any separable Bregman divergence $D_f$ that is
convex in its second argument.

\subsection{Universal PAC-Bayes Bounds}

Probably approximately correct (PAC)-Bayes theory blends Bayesian and
frequentist approaches to the analysis of machine learning.  The
PAC-Bayes formulation assumes a probability distribution on events
occurring in nature and a prior on the class of candidate hypotheses
(estimators) that express a learner's preference for some hypotheses
over others.  PAC-Bayes generalization bounds
\cite{mcallester1999pac,langford2005tutorial,mcallester2013pac} govern
the performance (loss) when stochastically selecting hypotheses from a
posterior distribution.  We begin this section with a summary of those
aspects of PAC-Bayes theory needed for our development.

Let $X$ be an explanatory variable\footnote{While the development
  generalizes naturally to collections of explanatory variables, to
  simplify the exposition we focus on a single such variable.}
(feature) and $Y$ an independent variable (target).  Assume that $X$
and $Y$ follow a joint probability distribution $p_{XY}$.  Let $\cH$
be a class of hypotheses (estimators) for $Y$, where each estimator
$q\in\cH$ is some functional of $X$.  As an example, in logistic
regression, each hypothesis is an estimator of the form
\eqref{eq:logistic} for some constants $\beta_0$ and $\beta_1$.

Next, we view $q$ as a realization of a random variable $Q$ that is
independent of $X$ and $Y$ and governed
by (prior) distribution $p_Q^0$ on $\cH$, and let
$l(y,q(x))$ be the loss between the realization $y$ and the estimate
$q(x)$, for a given estimator $q$ and loss function $l$, such that
$l_q(y,q(x)) \in [0, L_{\max}]$ for some constant $L_{\max}>0$ and all
$x$, $y$, and $q$.  We select $q\in\cH$ based on i.i.d.\ training
samples
\begin{equation*}
\cT_n\defeq\{(x_1,y_1),\dots,(x_n,y_n)\}
\end{equation*}
from $p_{X,Y}$ so as to minimize the
generalization loss
\begin{equation*}
L_q = \Ed{p_{X,Y}}{l(Y,q(X))}.
\end{equation*}
In particular, the selection is based on the training loss
\begin{equation*}
\hat{L}_q \defeq \frac1n \sum_{i=1}^n l(y_i, q(x_i)).
\end{equation*}

An example of a standard generalization bound of this type is the
following, in which $\cH$ is assumed to be countable.  
\begin{theorem}[PAC bound \cite{mcallester2013pac}]
\label{PAC}
Given training data $\cT_n$ from $p_{X,Y}$, with probability at least
$1-\delta$,
\begin{equation*} 
L_q \le \left(\frac{2\lambda}{2\lambda-1}\right)
\left(\hat{L}_q+\frac{\lambda L_{\max}}{n}
\log\frac{1}{\delta\, p_Q^0(q)} \right),
\end{equation*}
for all $q\in\cH$ and all $\lambda>1/2$.
\end{theorem}

In the PAC-Bayes extensions of \thmref{PAC}, we allow $\cH$ to be
continuous (uncountable).  Moreover, in addition to $p_Q^0$ we let $p_Q$
another distribution over $\cH$, and define
\begin{equation} 
L_{p_Q} \defeq \Ed{p_{X,Y}}{\tilde{L}_{p_Q}(X,Y)},
\end{equation}
and
\begin{equation}
\hat{L}_{p_Q} \defeq \frac1n \sum_{i=1}^n \tilde{L}_{p_Q}(x_i,y_i),
\label{eq:Lhat-pQ}
\end{equation}
where
\begin{equation} 
\tilde{L}_{p_Q}(x,y) \defeq \Ed{p_Q}{l(y,Q(x))}.
\label{eq:LpQ-xy}
\end{equation}

While tighter PAC-Bayes bounds have been developed---see, e.g.,
\cite{catoni2007pac,germain2009pac,langford2005tutorial,maurer2004note,seeger2002pac}---the
original is the following, which can be derived as a corollary of
results by Catoni \cite{catoni2007pac}.
\begin{theorem}[PAC-Bayes bound \cite{mcallester1999pac}]
\label{PAC-Bayes}
Given training data $\cT_n$ from $p_{X,Y}$, with probability
at least $1-\delta$,
\begin{equation*} 
L_{p_Q} \le\! \left(\frac{2\lambda}{2\lambda-1}\right)
\!\left(\hat{L}_{p_Q}\!+\!
\frac{\lambda L_{\max}}{n}\left(D_{\mathrm{KL}}(p_Q\|p_Q^0)\!+\!\log
\frac{1}{\delta}\right)\! \right),
\end{equation*}
for all $p_Q$ on $\cH$ and all $\lambda>1/2$.
\end{theorem}

Evidently, the bounds in both \thmref{PAC} and \thmref{PAC-Bayes} are
specific to the choice of loss function $l$.  For scenarios where such
a choice is not clear, a ``universal'' PAC-Bayes bound based on
log-loss, which we now develop, is useful.

A complication in the development is PAC-Bayes bounds apply only to
bounded loss functions as they focus on worst-case performance
\cite{mcallester2013pac}, and thus log-loss is inadmissible.
Different approaches have been introduced to overcome this limitation.
In \cite{mcallester2013pac}, McAllester suggests modifying an
unbounded loss by applying an ``outlier threshold'' $L_{\max}$ to
replace $l(y,q(x))$ with $\min\left\{l(y,q(x)), L_{\max}\right\}$,
which is always bounded..  This approach introduces analytical
difficulties as the new loss is typically neither continuous nor
convex.  

An alternative approach, which we follow and whose use is more
widespread, assumes that the underlying distribution for the data is
bounded away from zero \cite{haussler1992decision, bharadwaj2014pac,
  abe2001polynomial, shwartz2018representation}.  Equivalently, the
model $p_{Y|X}$ is not deterministic (singular), and the hypothesis
class is chosen accordingly.  Specifically, for some $\Delta>0$ we
have $p_{Y|X}(y|x),q(x) \in [\Delta,1-\Delta]$ for every $x$, $y$, and
$q$.  

Via the latter methodology, the loss function is bounded on the domain
of interest, and we obtain the following universal PAC-Bayes
inequality, a proof of which is provided in \appref{app:F}.
\begin{theorem}
\label{universal_PAC_Bayes}
Let $l(y,q)$ be a loss function that satisfies
\defref{def:admissible}, and $G$ its corresponding generalized entropy
function.  If $p(y|x),q(x) \in [\Delta,1-\Delta]$ for some $\Delta>0$
and every $x$, $y$, and $q\in\cH$, then with probability at least
$1-\delta$,
\begin{equation} 
L_{p_Q} \le 
\frac{2\lambda C(G)}{2\lambda-1}\!
\left(\hat{L}^{\log}_{p_Q} + \frac{\lambda L_{\max}}{n}\!
\left(D_{\mathrm{KL}}(p_Q\|p_Q^0) + \log \frac{1}{\delta}\right)
\right), 
\label{eq:PAC-Bayes-univ}
\end{equation}
for all $p_Q$ on $\cH$ and all $\lambda > 1/2$.  In \eqref{eq:PAC-Bayes-univ},
$L_{\max}=-\log\Delta$,
\begin{equation*}
C(G)> -\frac{1}{2}G''\left(\frac{1}{2}\right)
\end{equation*}
is a normalization constant that depends only on $G$, and
$\hat{L}^{\log}_{p_Q}$ is of the form \eqref{eq:Lhat-pQ}, where
$\tilde{L}_{p_Q}(x,y)$ is specialized to [cf.\ \eqref{eq:LpQ-xy}]
\begin{equation*} 
\tilde{L}^{\log}_{p_Q}(x,y) \defeq \Ed{p_Q}{l_{\log}(y,Q(x))},
\end{equation*}
with $l_{\log}$ as defined in \eqref{eq:logloss}.
\end{theorem}

\thmref{universal_PAC_Bayes} establishes that even when we do not know
\textit{a priori} the loss function with respect to which are to be
measured, it is often possible to bound the generalization loss.  Such
universal generalization bounds have potentially wide range of
applications.
 
\section{Discussion and Conclusions}
\label{conc}

In this work we introduce a fundamental inequality for two-class
classification problems.  We show that the KL divergence, associated
with the Bernoulli log-likelihood loss, upper bounds any divergence
measure that corresponds to a smooth, proper and convex binary loss
function.  This property makes the log-loss a universal choice, in the
sense that it controls any ``analytically convenient" alternative one
may be interested in.  This result has implications in a wide range of
applications.  There are many examples beyond those we have explicitly
described.  For instance, in binary classification trees
\cite{breiman1984classification}, the split criterion in each node is
typically chosen between the Gini impurity (which corresponds to
quadratic loss) and information-gain (which corresponds to log-loss).
The best choice for a splitting mechanism is a long standing open question
with many statistical and computational implications; see, e.g.,
\cite{painsky2017cross}.  Our results indicate that by minimizing the
information-gain we implicitly obtain guarantees for the Gini impurity
(but not vice-versa).  This provides a new and potentially useful
perspective on the question.

Finally, by viewing our bounds from a Bregman divergence perspective,
we extend the well-studied $f$-divergence inequalities by providing
complementary Bregman inequalities.  Collectively, these results
contribute to our growing understanding understanding of the
fundamental role that KL divergence plays in these two important
classes of divergences.

\appendices

\section{Bregman Divergence Characterization}
\label{app:A}

Following \cite{broniatowski2019some}, a Bregman divergence
generator is a continuous, strictly convex (finite)
function $f\colon\cS\mapsto\reals$ on some appropriately chosen open
interval $\cS=(a,b)$ such that $[a,b]$ covers (at least) the union of
the ranges of $s$ and $t$, as appears in
\eqref{eq:bregdiv-bin}---e.g., $\cS=[0,1]$ in the binary
classification problem of \secref{main result}.
Due to condition \ref{II}, we further restrict our
attention to continuously differentiable $f$.  

We continuously extend $f$ to $\bar{f}\colon[a,
  b]\mapsto\reals\cup\{+\infty\}$ via
\begin{equation*}
\bar{f}(s) \defeq 
\begin{cases} 
\ds\lim_{s\to a} f(s) & s=a \\
f(s) & s\in(a,b) \\
\ds\lim_{s\to b} f(s)& s=b,
\end{cases}
\end{equation*}
which can be infinite only for $s\in\{a,b\}$.  Moreover, we
continuously extend the derivative
$f'(s)\colon(a,b)\mapsto\reals$ to
$\bar{f}'\colon[a, b] \mapsto\reals\cup\{-\infty,+\infty\}$ via
\begin{equation*} 
\bar{f}'(s)\defeq \begin{cases}
\ds\lim_{s\to a}f'(s) & s=a \\
f'(s) & s\in(a,b) \\
\ds\lim_{s\to b}f'(s) & s=b.
\end{cases}
\end{equation*}

Using these extensions, for $\cS=[a,b]$, we let
\begin{equation*}
D_f(s\|t)\defeq\bar{\psi}(s,t),
\end{equation*}
where
$\bar{\psi}_f\colon[a,b]^2\mapsto\reals\cup\{+\infty\}$ is following lower
semi-continuous nonnegative function.
First,
\begin{equation*}
\bar{\psi}_f(s,t) \defeq 
\bar{f}(s)-f(t)-(s-t)f'(t),\quad s\in[a,b],\ t\in (a,b).
\end{equation*}
Next, for $s\in(a,b)$,
\begin{align*}
&\bar{\psi}_f(s,a) \notag\\
&\ \defeq \begin{cases}
\ds f(s)-s\bar{f}'(a) +\lim_{t\to a} \left[ t\bar{f}'(a)-f(t)\right]
  & \bar{f}'(a)>-\infty \\
\infty  &\bar{f}'(a)=-\infty 
  \end{cases}
\end{align*}
and
\begin{align*} 
&\bar{\psi}_f(s,b) \notag\\
&\ \defeq
\begin{cases}
\ds f(s)-s \bar{f}'(b) +\lim_{t\to b} \left[ t\bar{f}'(b)-f(t) \right] &
\bar{f}'(b)<+\infty \\
\infty & \bar{f}'(b) = +\infty, 
\end{cases}
\end{align*}
where we note the limits exist but may be infinite. Finally,
\begin{equation*} 
\bar{\psi}_f(s,t) = \begin{cases}
0 & (s,t)=(a,a) \\
\begin{aligned} \ds\smash{\lim_{s\to a}}\vphantom{\lim} &\left[
    f(s)-s\bar{f}'(b)\right]  \\ 
&{}+\lim_{t\to b} \left[ t\bar{f}'(b)-f(t)\right]\vphantom{\lim_{\substack{a\\b}}} 
\end{aligned} & (s,t)=(a,b)\\
\begin{aligned} \ds\smash{\lim_{s\to b}}\vphantom{\lim} &\left[ f(s)-s\bar{f}'(a)\right] \\
&{}+\lim_{t\to a} \left[ t\bar{f}'(a)-f(t)\right]
\end{aligned} & (s,t)=(b,a)\\
0 & (s,t)=(b,b). \\
\end{cases}
\end{equation*}

\section{Weight Functions of Smooth Proper Losses}
\label{app:extra}

As a complementary view of weight functions, we note that when a
smooth loss function is proper, its expected loss satisfies
\begin{equation*} 
\left.\frac{\p}{\p q} L(p,q)\right|_{q=p} = p\, l'_1(p) + (1-p)\,
l'_0(p) = 0,
\end{equation*}
whence
\begin{equation}
\frac{-l'_1(p)}{1-p}=\frac{l'_0(p)}{p} = w(p),
\label{eq:wf-alt}
\end{equation}
where the last equality in \eqref{eq:wf-alt} is obtained by matching
terms in the forms \eqref{eq:L-def} and \eqref{eq:L-alt}, and using
\eqref{eq:wf-def}.  Shuford \textit{et
  al.}\ \cite{shuford1966admissible} establish that the converse is
also true: a smooth loss function is proper only if \eqref{eq:wf-alt}
holds for some nonnegative $w(p)$ that satisfies
$\int_\epsilon^{1-\epsilon} w(p)\, \diff p < \infty$, for all
$\epsilon>0$.

\section{Proof of \thmref{main_theorem}} 
\label{app:B}

First, due to the convexity of the loss (with respect to $q$), we have
\begin{equation}
\frac{\p^2}{\p q^2} L(p,q) = \frac{\p}{\p
  q}w(q)(q-p) =w(q)+(q-p)w'(q)\ge 0 
\label{eq:L-curve}
\end{equation}
for every fixed $p\in[0,1]$ and $q\in (0,1)$.  Specializing
\eqref{eq:L-curve} to the cases $p=0$ and $p=1$ then yields
\begin{equation}
-\frac{1}{q}\le \frac{w'(q)}{w(q)}\le\frac{1}{1-q}
\label{eq:wpw-rel}
\end{equation}
for all $q\in (0,1)$.   
In turn, \eqref{eq:wpw-rel} implies
\begin{align*}
-\int_0^{1/2} \frac{1}{q}\,\diff q &\le \int_0^{1/2}
\frac{w'(q)}{w(q)}\,\diff q \le \int_0^{1/2} \frac{1}{1-q}\,\diff q \\
-\int_{1/2}^1 \frac{1}{q}\,\diff q &\le \int_{1/2}^1
\frac{w'(q)}{w(q)}\,\diff q \le \int_{1/2}^1 \frac{1}{1-q}\,\diff q,
\end{align*}
i.e.,
\begin{subequations} 
\begin{align}
\frac{w(1/2)}{2(1-q)}\le w(q) \le \frac{w(1/2)}{2q},\qquad q\in(0,1/2)
\label{basic_inq2}\\
\frac{w(1/2)}{2q}\le w(q) \le
\frac{w(1/2)}{2(1-q)},\qquad q\in[1/2,1). \label{basic_inq1}
\end{align}
\end{subequations}
Similar results appear in, e.g., \cite[Theorem 29]{reid2010composite}.
We emphasize that we have not assumed that $w(\cdot)$ is integrable on
$(0, 1)$, so as to accommodate loss functions such that $l_0(\cdot)$
and/or $l_1(\cdot)$ are unbounded at $0$ and $1$, respectively\cite{buja2005loss}.

Next, we show there exists a constant $C$ such that 
\begin{equation} 
R(p,q) \defeq C \, D_{\mathrm{KL}}(p\|q)-D_{-G}(p\|q)
\label{eq:Rpq-def}
\end{equation}
is nonnegative for all $p,q \in [0,1]$.  For any
$p\in[0,1]$, since $R(p,p)=0$ it suffices to show that 
$R(p,\cdot)$ has a minimum at $p$ for a suitable choice of $C$.
From
\begin{equation}
\frac{\p}{\p q}
R(p,q)=(q-p)\left(\frac{C}{q(1-q)}-w(q)\right),
\label{fd} 
\end{equation}
we see that $q=p$ is a unique stationary point.   Moreover, this
stationary point is a minimum when
\begin{align} 
&\frac{\p^2}{\p q^2} R(p,q)\Biggr|_{q=p} \notag\\
&\quad= \left[ C\left(\frac{p}{q^2}+\frac{1\!-\!p}{(1\!-\!q)^2}\right)-w(q)-(q\!-\!p) \,w'(q)\right] \Biggr|_{q=p} \notag\\
&\quad= \frac{C}{p(1-p)} - w(p) >0,
\label{sd}
\end{align}
for all $p\in(0,1)$.

Now for every $q\in(0,1/2)$, we have
\begin{subequations} 
\begin{equation} 
\frac{C}{q(1-q)}-w(q) > \frac{C}{q}-w(q) 
\ge \frac{1}{q}\left(C-\frac{1}{2}w\left(\frac{1}{2}\right)\right),
\label{proof1}
\end{equation}
where the first inequality follows since $q>0$, and the last inequality
follows from \eqref{basic_inq2}.  Similarly, for $q\in[1/2,1)$ we have
\begin{equation} 
\frac{C}{q(1-q)}-w(q) > \frac{C}{1-q}-w(q) \ge
\frac{1}{1-q}\left(C-\frac{1}{2}w\left(\frac{1}{2}\right)\right),
\label{proof2}
\end{equation}
\label{eq:proof}%
\end{subequations}
where the first inequality follows since $q<1$, and the last inequality
follows from \eqref{basic_inq1}.  Hence, choosing
\begin{equation} 
C>\frac{1}{2}
\,w\left(\frac{1}{2}\right)=-\frac{1}{2}\,G''\left(\frac12\right) 
\label{eq:C-suff}
\end{equation}
ensures the right-hand side of (the relevant variant of)
\eqref{eq:proof} is positive for all $q\in(0,1)$,
and thus \eqref{sd} holds for all $p\in(0,1)$.   Hence, we conclude
that $R(p,q)\ge0$ for all $p,q\in(0,1)$.

Next consider the case $p\in\{0,1\}$ and $q\in(0,1)$.  If choose $C$
according to \eqref{eq:C-suff}, then \eqref{sd} holds for all
$p\in(0,1)$.  In this case, \eqref{fd} must be strictly positive for
all $q\in(0,1)$ when $p=0$, so $R(0,\cdot)$ is monotonically increasing, and thus its minimum is attained at $0$.  Likewise \eqref{fd}
must be strictly negative for all $q\in(0,1)$ when $p=1$, so
$R(1,\cdot)$ is monotonically decreasing, and thus its minimum is
attained at $1$.  In turn, since $R(0,0)=R(1,1)=0$, it follows that
$R(p,q) \ge 0$ also holds for $p \in \{0,1\}$ and $q\in(0,1)$.

It remains to consider the case $p\in[0,1]$ and $q\in\{0,1\}$.  When
$p=q\in\{0,1\}$ we have $R(p,q)\ge 0$ since $R(0,0)=R(1,1)=0$.
Finally, when $p \ne q\in\{0,1\}$ we have $D_{\mathrm{KL}}(p\|q)$ is
unbounded, so \eqref{main_ineq} holds trivially.\hfill\IEEEQED

\section{Proof of \thmref{divergence_theorem_general}} 
\label{app:C}

It suffices to show that
\begin{equation*}
R(p^m,q^m) \defeq C(f)\, \tilde{D}_{\mathrm{KL}}(p^m\|q^m)-D_f(p^m\|q^m)
\end{equation*}
is nonnegative for all $p^m,q^m\in [0,1]^m$.  Using
\eqref{eq:bregdiv-m} in the form
\begin{equation} 
D_f(p^m\|q^m) 
=f(p^m)-f(q^m)-\sum_{k=1}^m \frac{\p}{\p q_k} (p_k-q_k)f(q^m),
\end{equation}
we have
\begin{align}
\frac{\p}{\p p_i} R(p^m,q^m) 
&= \left(C(f)\log p_i - \frac{\p f(p^m)}{\p p_i}\right)\notag\\
&\qquad\qquad{}-\left(C(f)\log q_i - \frac{\p f(q^m)}{\p
  q_i}\right) \label{grad}
\end{align}
and, in turn,
\begin{align} 
\frac{\p^2}{\p p_i\p p_j} R(p^m,q^m) 
&= C(f)\,\frac{\kron\{i=j\}}{p_i} - \frac{\p^2 f(p^m)}{\p p_i\p p_j}\notag\\
&= \left[ C(f)\,H_\mathrm{KL}(p^m) - H_f(p^m)\right]_{i,j},
\label{eq:R-hess}
\end{align}
where $[\cdot]_{i,j}$ denotes the $i,j$\/th element of its matrix
argument.  Hence, it follows that $R(\cdot,q^m)$ is strictly convex if
there exists a constant $C(f)$ such that \eqref{pd_condition} is
satisfied.  Moreover, from \eqref{grad} we have that $p^m=q^m$ is a
stationary point, so provided \eqref{pd_condition} is
statisfied, this stationary point is a minimum.  Finally, since
$R(p^m,p^m)=0$, it follows that $R(p^m,q^m)\ge0$ for all
$p^m,q^m\in[0,1]^m$.\hfill\IEEEQED

\section{Proof of \protect\corolref{corol:mahalanobis}}
\label{app:D}

First, let
\begin{equation*} 
V(p^m) \defeq C(Q)\, H_{\mathrm{KL}}(p^m) - Q, \quad 
p^m \in [0,1]^m,
\end{equation*}
with
\begin{equation*} 
\lambda_1(V) \ge \dots \ge \lambda_m(V) \defeq \lambda_{\min}(V) 
\end{equation*}
denoting its eigenvalues, and note that according to
\eqref{pd_condition} it suffices to show that when $C(Q)$ satisfies
\eqref{eq:mahalanobis-cond}, we have $V(p^m)\succ 0$, for which the condition
$\lambda_{\min}(V) >0$ is equivalent.

Next, let $\Vt=V(1^m)$, whose eigenvalues we denote via
\begin{equation*} 
\lambda_1(\Vt) \ge \dots \ge \lambda_m(\Vt) \defeq \lambda_{\min}(\Vt),
\end{equation*}
and note that for every $x^m\in\reals^m$, 
\begin{align*} 
(x^m)^\T V x^m
&= C(Q) \sum_{i=1}^m \frac{x_i^2}{p_i} - (x^m)^\T Q\, x^m \\
&\ge  C(Q) \sum_{i=1}^m x_i^2 - (x^m)^\T Q\, x^m \\
&= (x^m)^\T \Vt x^m.
\end{align*}
Hence, \begin{align*} 
\lambda_{\min}(V) 
&= \min_{\{x^m\colon \sum_i x_i^2=1\}} (x^m)^\T V x^m \\
&\ge \min_{\{x^m\colon \sum_i x_i^2=1\}} (x^m)^\T V x^m = \lambda_{\min}(\Vt),
\end{align*}
where the equalities follow from the Rayleigh quotient theorem
\cite[Theorem~4.2.2]{hj12}.

Finally, $\lambda_i(\Vt) = C(Q)-\lambda_i(Q)$ since\footnote{We use
  $I$ to denote the identity matrix.} $\Vt=C(Q)\, I - Q$, so 
\begin{equation*} 
\lambda_{\min}(V) \ge \lambda_{\min}(\Vt) = C(Q)
-\lambda_{\max}(Q).
\end{equation*}
Accordingly, setting 
$C(Q)>\lambda_{\max}(Q)$ yields $V(p^m) \succ 0$ for all
$p^m\in[0,1]^m$.\hfill\IEEEQED

\section{Proof of \thmref{divergence_theorem}} 
\label{app:E}

First, note that 
\begin{subequations}
\begin{align}
\frac{\p}{\p q} d_g(p\|q) &= (q-p)g''(q) \label{eq:dg-1}\\
\frac{\p^2}{\p q^2} d_g(p\|q) &= (q-p)g'''(q) + g''(q).
\label{eq:dg-2}
\end{align}
\label{eq:dg-derivs}%
\end{subequations}
Since $d_g(p,\cdot)$ is convex for every $p\in[0,1]$,
\eqref{eq:dg-2} is nonnegative for every $p\in[0,1]$ and $q \in
(0,1)$.  Choosing $p=0$
we obtain 
\begin{equation*} 
-\frac{1}{q} \le \frac{g'''(q)}{g''(q)},
\qquad q\in(0,1). 
\end{equation*}
Hence, 
we have
\begin{equation*} 
-\int_{q}^1 \frac1{q}\,\diff q \le \int_{q}^1
\frac{g'''(q)}{g''(q)} \, \diff q,
\end{equation*}
whence
\begin{equation} 
   g''(q)\le \frac{g''(1)}{q}.
\label{appendix_C_inequality}
\end{equation}

Next, following an approach similar to that in the proof of
\thmref{main_theorem}, we define 
\begin{subequations} 
\begin{equation} 
R(p^m,q^m) \defeq \sum_{i=1}^m r(p_i,q_i)
\label{eq:Rm-def}
\end{equation}
with
\begin{equation} 
r(p,q) \defeq C\, \tilde{d}_{\mathrm{KL}}(p\|q)-d_g(p\|q),
\label{eq:r-def}
\end{equation}
\end{subequations}
and show that when $C$ is chosen as prescribed, \eqref{eq:Rm-def} is
nonnegative for $p^m,q^m\in[0,1]^m$. Note that it is sufficient to
show that for such $C$, \eqref{eq:r-def} is nonnegative for every
$p,q\in[0,1]$.  

Accordingly, we fix $p$ and 
analyze $r(p,q)$ with respect to $q$.  Via \eqref{eq:dg-derivs}
(including its specialization to \eqref{eq:KL-g}) we obtain
\begin{subequations} 
\begin{align} 
\frac{\p}{\p q}
r(p,q) &= (q-p)\left(\frac{C}{q}-g''(q) \right) \label{fd2} \\
\frac{\p^2}{\p q^2} r(p,q)
&=
C\,\frac{p}{q^2_i}-g''(q)-(q-p)g'''(q). \label{sd2}   
\end{align}
\end{subequations}

First, consider the case $p \in (0,1)$.  Since $r(p,p)=0$, if
$r(p,q)\ge0$ then a global minimum of $r(p,\cdot)$ must occur at
$p$.  Proceeding, from \eqref{fd2}, we see that the unique
stationary point is $q=p$.  Moreover, this stationary point is a
minimum when 
\begin{equation*}
\frac{\p^2}{\p q^2} r(p,q) \Biggr|_{q=p} =
\frac{C}{p} - g''(p) 
\end{equation*}
is positive, from which 
we obtain the requirement
\begin{equation}
\label{inq2}
\frac{C}{q}-g''(q)>0, \quad \text{for all $q\in(0,1)$}.
\end{equation}
Choosing $C>g''(1)$ we obtain
\begin{equation*}
\frac{C}{q} - g''(q) >
\frac{g''(1)}{q} - g''(q) \ge 0,
\end{equation*}
where the last inequality follows from \eqref{appendix_C_inequality}.
Hence, $r(p,q)\ge0$ for $p,q\in(0,1)$.

Next, consider the case $p\in\{0,1\}$.  Again, with the choice
$C>g''(1)$, \eqref{inq2} holds for all $q$, and thus \eqref{fd2} is
positive for $q\in(0,1)$ when $p=0$, so $r(0,\cdot)$ is an
increasing function.  Since $r(0,0)=0$, then, we conclude
$r(0,q)\ge0$.  Likewise, thus \eqref{fd2} is negative for
$q\in(0,1)$ when $p=1$, so $r(1,\cdot)$ is a decreasing function.
Since $r(1,1)=0$, then, we conclude $r(1,q)\ge0$.  Hence,
$r(p,q)\ge0$ for $p\in\{0,1\}$ and $q\in(0,1)$.

It remains only to consider the case $q\in\{0,1\}$, for any
$p\in[0,1]$.  When $p=q$, we have $r(p,q)=r(q,q)=0$.  When $p\ne q=0$,
\eqref{eq:dKL-def} is unbounded so $r(p,0)\ge0$.  For the case $q=1$,
straightforward calculation yields
\begin{equation}
\frac{\p}{\p p} r(p,1) = \alpha(p)-\alpha(1),
\quad\text{with}\ \alpha(p) \defeq C \log p - g'(p).
\label{eq:p-deriv}
\end{equation}
But 
\begin{equation*}
\alpha'(p) = \frac{C}{p} - g''(p),
\end{equation*}
which matches the left-hand side of \eqref{inq2}, and thus is positive
for all $p\in(0,1)$ when $C>g''(1)$, in which case $\alpha(\cdot)$ is
an increasing function.  As a result, \eqref{eq:p-deriv} is negative
for $p\in(0,1)$, and thus $r(\cdot,1)$ is a decreasing function.
Since, in addition, $r(1,1)=0$, we conclude $r(p,1)\ge0$. Hence,
$r(p,q)\ge0$ for $p\in[0,1]$ and $q\in\{0,1\}$.\hfill\IEEEQED

\section{Proof of \thmref{universal_PAC_Bayes}} 
\label{app:F}

The following lemma will be useful.
\begin{lemma}
\label{lemma1}
If $l(y,q)$ is a loss function that satisfies 
\defref{def:admissible}, with corresponding generalized entropy function
$G$, then 
\begin{equation*}
G(p)-CG_{\log}(p) \le 0,\quad\text{for all $p,q\in[0,1]$},
\end{equation*}
when 
\begin{equation}
C>-\frac12\, G''\left(\frac12\right),
\label{eq:C-cond}
\end{equation}
where $G_{\log}(p)$ is the Shannon entropy as defined in
\eqref{eq:shannon-ent}.
\end{lemma}
\begin{IEEEproof}
With
\begin{equation*}
R(p)=G(p)-CG_{\log}(p)
\end{equation*}
we have
\begin{subequations}
\begin{align} 
\frac{\p}{\p p} R(p) &=
G'(p)-C\log\frac{1-p}{p} \label{fd1} \\
\frac{\p^2}{\p p^2} R(p) &=
G''(p)+\frac{C}{p(1-p)}=\frac{C}{p(1-p)}-w(p), \label{sd1} 
\end{align}
where to obtain the second equality in \eqref{sd1} we have used
\eqref{eq:wf-def}.
\end{subequations}
In turn, using \eqref{eq:proof} from the proof of
\thmref{main_theorem}, we likewise conclude that choosing $C$ according to
\eqref{eq:C-cond} ensures that
\begin{equation*}
\frac{C}{p(1-p)} - w(p) > 0,
\end{equation*}
in which case $R(p)$ is strictly convex.  In addition, we have
\begin{equation*}
G(p) = L(p,p) = (1-p)\, l_0(p) + p\, l_1(p),
\end{equation*}
where the first and second qualities follow from \eqref{eq:L-alt} and
\eqref{eq:L-def}, respectively, and thus using \eqref{eq:regular} we
have $G(p)=0$ for $p\in\{0,1\}$.  Since $G_{\log}(p)=0$ for
$\{0,1\}$ as a special case, it follows that $R(p)=0$
for $p\in\{0,1\}$.   Hence, $R(p)\le0$.
\end{IEEEproof}

Proceeding to the proof of \thmref{universal_PAC_Bayes}, from
\eqref{eq:regret-alt} with \eqref{eq:regret} we obtain
$D_{-G}(p\|q)=L(p,q)-G(p)$, which when used in conjunction with
\eqref{main_ineq} of \thmref{main_theorem} yields
\begin{align} 
L(p,q) 
&\le C\, L_{\log}(p,q) + G(p) - C\,G_{\log}(p) \label{ineqApp}\\
&\le C\, L_{\log}(p,q), \label{ineqApp2}
\end{align}
where in \eqref{ineqApp}
\begin{equation*}
L_{\log}(p,q) \defeq \E{l_{\log}(Y,q)},
\end{equation*}
and where to obtain \eqref{ineqApp2} we have
used \lemref{lemma1}.

Next, we have
\begin{align}
L_{p_Q} &=\Ed{p_{X,Y}}{\Ed{p_Q}{ l(Y,Q(X))}} \notag\\
&= \Ed{p_X p_Q}{\Ed{p_{Y|X}(\cdot|X)}{l(Y,Q(X))}} \label{eq:switch} \\
&\le \Ed{p_X p_Q}{C\,
  \Ed{p_{Y|X}(\cdot|X)}{l_{\log}(Y,Q(X))}} \label{eq:inner-bound} \\
&=C\,L^{\log}_{p_Q}, \label{eq:LpQ-bound}
\end{align}
where to obtain \eqref{eq:inner-bound} we have used an instance of
\eqref{ineqApp2} to bound the inner expectation in \eqref{eq:switch}.

Moreover, since $p_{Y|X}(y|x),q(x) \in [\Delta, 1-\Delta]$ we have
\begin{equation}
l_{\log}(Y,Q(X)) \in [0,-\log\Delta]
\label{ll-bound}
\end{equation}
with probability one.

Finally, using \eqref{eq:LpQ-bound} followed by \thmref{PAC-Bayes}
specialized to the log-loss, together with \eqref{ll-bound}, we obtain
that with probability $1-\delta$,
\begin{align*} 
L_{p_Q} &\le C\, L^{\log}_{p_Q} \notag\\
&\le \frac{2\lambda C}{2\lambda1-1}\! \left(\hat{L}_{p_Q}^{\log}+
\frac{\lambda L_{\max}}{n}\! \left(D_{\mathrm{KL}}(p_Q\|p_Q^0) +
\log\frac1\delta\right) \right), 
\end{align*}
for any $\lambda>1/2$ and $L_{\max}=-\log\Delta$.\hfill\IEEEQED

\bibliographystyle{IEEEtran}
\bibliography{refs}

\begin{IEEEbiographynophoto}{Amichai Painsky}  (S’12–M’18)
received his B.Sc.  in Electrical Engineering from Tel Aviv University
(2007), his M.Eng.  degree in Electrical Engineering from Princeton
University (2009) and his Ph.D.  in Statistics from the School of
Mathematical Sciences in Tel Aviv University.  He was a Post-Doctoral
Fellow, co-affiliated with the Israeli Center of Research Excellence
in Algorithms (I-CORE) at the Hebrew University of Jerusalem, and the
Signals, Information and Algorithms (SIA) Lab at MIT
(2016-2018).  Since 2019, he is a faculty member at the Industrial
Engineering Department at Tel Aviv University, where he leads the
Statistics and Data Science Laboratory.   His research interests
include Data Mining, Machine Learning, Statistical Learning and
Inference, and their connection to Information Theory.
\end{IEEEbiographynophoto}

\begin{IEEEbiographynophoto}{Gregory W.~Wornell} (S'83-M'91-SM'00-F'04)
received the B.A.Sc.\ degree from the University of British Columbia,
Canada, and the S.M. and Ph.D. degrees from the Massachusetts
Institute of Technology, all in electrical engineering and computer
science, in 1985, 1987 and 1991, respectively.

Since 1991 he has been on the faculty at MIT, where he is the Sumitomo
Professor of Engineering in the Department of Electrical Engineering
and Computer Science.  At MIT he leads the Signals, Information, and
Algorithms Laboratory within the Research Laboratory of Electronics.
He is also chair of Graduate Area I (information and system science,
electronic and photonic systems, physical science and nanotechnology,
and bioelectrical science and engineering) within the EECS
department's doctoral program.  He has held visiting appointments at
the former AT\&T Bell Laboratories, Murray Hill, NJ, the University of
California, Berkeley, CA, and Hewlett-Packard Laboratories, Palo Alto,
CA.

His research interests and publications span the areas of information
theory, statistical inference, signal processing, digital
communication, and information security, and include architectures for
sensing, learning, computing, communication, and storage; systems for
computational imaging, vision, and perception; aspects of
computational biology and neuroscience; and the design of wireless
networks.  He has been involved in the Information Theory and Signal
Processing societies of the IEEE in a variety of capacities, and
maintains a number of close industrial relationships and activities.
He has won a number of awards for both his research and teaching,
including the 2019 IEEE Leon K. Kirchmayer Graduate Teaching Award.
\end{IEEEbiographynophoto}
\vfill

\end{document}